\documentclass[aps,pre,preprint,showpacs]{revtex4}
\usepackage{amsmath}
\usepackage{times}

\begin{document}

\title{Kinetic Theory and Hydrodynamics of Dense, Reacting Fluids far from
Equilibrium}
\author{James F. Lutsko}
\email{jlutsko@ulb.ac.be}
\affiliation{Center for Nonlinear Phenomena and Complex Systems\\
Universit\'{e} Libre de Bruxelles\\
Campus Plaine, CP 231, 1050 Bruxelles, Belgium}
\date{\today }

\begin{abstract}
The kinetic theory for a fluid of hard spheres which undergo endothermic
and/or exothermic reactions with mass transfer is developed. The exact
balance equations for concentration, density, velocity and temperature are
derived. The Enskog approximation is discussed and used as the basis for the
derivation, via the Chapman-Enskog procedure, of the Navier-Stokes-reaction
equations under various assumptions about the speed of the chemical
reactions. It is shown that the phenomenological description consisting of a
reaction-diffusion equation with a convective coupling to the Navier-Stokes
equations is of limited applicability.
\end{abstract}

\maketitle

\section*{Introduction}

An understanding of chemically reactive flows is necessary in a wide range
of disciplines including astrophysics, plasma physics and the chemical
industry. Recently, applications in aerospace engineering have lead to a
number of studies aimed at deriving phenomenological equations, the
Navier-Stokes equations coupled to reactions, from the Boltzmann equation
for increasingly complex systems including internal degrees of freedom and
three-body interactions\cite{Chikhaoui94, Chikhaoui00}. However, these
investigations do not exhaust the range of interesting applications. A
number of important applications arise in the physio-chemistry of cavitating
bubbles\cite{Bubbles}. Aside from obvious examples such as the study of
flames and explosions, a recent area of interest is sonochemistry in which
ultrasound is used to induce conditions of extreme temperature and pressure
inside bubbles with the effect of dramatically increasing the rates of
chemical reactions\cite{SomnoChem}. Closely related is the phenomena of
somnoluminescence - in which a fluid irradiated with ultrasound emits light
- is believed to be caused by pressure waves acting on small bubbles of gas
in the fluid(see, e.g. \cite{SomnoReview}). The bubbles are subjected to
such rapid compression that shockwaves may develop and the concentrated
energy drives many chemical reactions particularly when the shocks reach the
center of the bubbles giving rise to high temperatures and densities. In
fact, it has been suggested\cite{Didenko2002} that some (endothermic)
reactions may play an important role in limiting the temperatures reached in
the center of the bubble. It is therefore of interest for these
applications, as well as some of the others mentioned above, to understand
the phenomenological equations governing a reacting gas under extreme
conditions and far from equilibrium. The Boltzmann equation cannot be
considered an adequate basis for such a study due to the fact that it is
only applicable at asymptotically low densities. In fact, the only simple
fluid that is amenable to analytic investigation at finite densities is one
composed of hard spheres. The purpose of this paper is therefore to review
the kinetic theory of reacting hard-sphere systems and to use this as a
basis for a hydrodynamic description of a reacting fluid far from
equilibrium. In particular, the kinetic theory will be used to derive the
exact balance equations describing the local concentration, density,
velocity and temperature fields from which the extension of the
Navier-Stokes equations to include reactions is developed based on the
Chapman-Enskog procedure applied to the Enskog approximation to the kinetic
theory. A primary result will be to show that the usual phenomenological
description consisting of the Navier-Stokes equations coupled to a
reaction-diffusion-advection equation is only applicable if the chemical
reactions take place on a time scale which is comparable to the dissipative
time scale $\lambda k^{2}$, where $\lambda $ is a transport coefficient and $%
k$ a typical wavevector. If the reactions are slower, then all hydrodynamic
relaxation takes place before the chemistry gets started and chemistry and
hydrodynamics are effectively decoupled. Faster reactions, with time-scale
comparable to $ck$, where $c$ is the sound velocity, leads to additional
couplings of the reactions to the hydrodynamic fields. Even faster reactions
lead to the chemistry taking place so fast that hydrodynamics is irrelevant.

The hard-sphere interaction model has proven remarkably useful as models of
single- and multi-component simple fluids since, in many respects, the
phenomenology of the hard-sphere systems and atoms interacting via more
realistic pair potentials is qualitatively identical. For example,
hard-sphere systems exhibit the full range of transport coefficients found
in all simple fluids\cite{McLennan}, possess a freezing transition\cite%
{Evans} and the structure of hard-sphere fluids in equilibrium is not much
different from that of any other fluid\cite{HansenMcdonald}. On the
theoretical side, the equation of state of hard-sphere fluids is easily
modeled\cite{HansenMcdonald}, kinetic theory is simplified by the fact that
only binary collisions are important and it is possible to formulate an
extension of the Boltzmann equation - the so-called Revised Enskog Theory or
RET\cite{RET}- which not only describes the transport properties of
multi-component hard-sphere fluids at finite densities but which also
describes transport in the solid state\cite{KDEP}. More recently, inelastic
hard spheres have been used as a model for driven granular fluids with
similar success. The hard-sphere interaction is therefore an ideal model for
understanding the extreme conditions occuring in sonochemical experiments.

The kinetic theory of chemically reacting hard spheres has in fact been
discussed in the literature\cite{Kapral78, Bose, Cukier, KapralAdvChem}. The
principal aim of these studies was to investigate contributions to the
reaction rates coming from dense fluid effects (e.g., ring kinetic theory
leading to mode-coupling models) at equilibrium. In these studies, the atoms
carry labels indicating their species (sometimes called their color) and all
intrinsic properties like the atomic radius and mass is specific to the
species. When the atoms collide, there is a probability that a reaction
takes place in which the species labels, and hence atomic properties,
change. The probability typically depends on the rest-frame energy of the
colliding atoms: if the rest frame energy is greater than some specified
activation energy, the reaction can take place with a probability that is,
in general, a function of the relative energy. Energy may be gained or lost
(exo- or endothermic reactions) but the sizes of the atoms are generally
invariant since, were they to also vary, a collision could result in one of
the atoms overlapping with third atom. (Technically, there is no reason that
atoms could not get \emph{smaller} and most results would apply to such a
model). Besides being restricted to a chemistry consisting of color labels
(and so excluding, e.g., the exchange of mass upon collision), a common
assumption in earlier work is that the chemical reactions are \emph{slow}
compared to other transport processes. Since the rate of chemical reactions
is generally determined by the ratio of the temperature to the activation
energies and the difference in the concentrations, this implies that the
results are only applicable near equilibrium or for low temperatures. One of
the primary goals of the present work is to indicate how the
phenomenological description (Navier-Stokes equations coupled to a set of
advective-reaction-diffusion equations) must be modified to account for
large deviations from equilibrium.

The organization of this paper is as follows. The second Section develops
the formal statistical description of the system. The discussion of possible
collision rules is followed by the development of the Liouville equation and
the exact balance laws describing the evolution of the local mass, energy,
momentum and partial densities. The Enskog approximation is also introduced.
The third Section discusses the Chapman-Enskog solution of the Enskog
equation and, particularly, the difference between the assumptions of fast
and slow chemical reactions. It is shown that, if the chemical reactions are
sufficiently slow, the fluid may be described by the Navier-Stokes equations
for the total mass, energy and momentum densities and a
reaction-diffusion-convection equation for the concentrations with the only
coupling between the two being the convective term occurring in the latter
(i.e., the ''usual'' description). However, under less restrictive
assumptions, the reactions are shown to depend in a much more complicated
way on the hydrodynamic fields. The paper concludes with a discussion of the
physical meaning of the different assumptions.

\section{Statistical Mechanics of Reacting Hard-Spheres}

Consider a system of $N$ hard spheres of various species confined to a
volume $V$ with positions $\left\{ \overrightarrow{q}_{i}\right\} _{i=1}^{N}$
and momenta $\left\{ \overrightarrow{p}_{i}\right\} _{i=1}^{N}$. Each atom
will also be described by a set of discrete labels, $l_{i}$ for the $i$-th
atom, which fix its ''chemical'' properties (mass, hard-sphere diameter and
reaction parameters). When two atoms collide, both the mechanical variables
and the species labels of the atoms change. The dynamical variables are
altered according to some deterministic collision rule so that for
collisions between the $i$-th and $j-$th atoms, 
\begin{eqnarray}
x_{i} &\rightarrow &x_{i}^{\prime }=\widehat{b}_{l_{i}l_{j}}^{l_{i}^{\prime
}l_{j}^{\prime }}x_{i} \\
x_{j} &\rightarrow &x_{j}^{\prime }=\widehat{b}_{l_{i}l_{j}}^{l_{i}^{\prime
}l_{j}^{\prime }}x_{j},  \notag
\end{eqnarray}%
where the collision operator $\widehat{b}_{l_{i}l_{j}}^{l_{i}^{\prime
}l_{j}^{\prime }}$ describes a collision involving the reaction $%
l_{i}+l_{j}\rightarrow l_{i}^{\prime }+l_{j}^{\prime }$ (i.e., the $i$-th
atom changes from species $l_{i}$ to $l_{i}^{\prime }$, etc.) and clearly,
one expects that $\widehat{b}_{l_{i}l_{j}}^{l_{i}^{\prime }l_{j}^{\prime }}=%
\widehat{b}_{l_{j}l_{i}}^{l_{j}^{\prime }l_{i}^{\prime }}$. Because the
species change instantaneously and randomly upon collision, the species
labels must be viewed as discrete random variables. Attention here will be
restricted to the model in which the probability of making a particular
transition is given by some function of the relative phases of the two
colliding atoms $K_{l_{i}l_{j}}^{l_{i}^{\prime }l_{j}^{\prime }}\left(
x_{ij}\right) $ where the notation indicates that this probability depends
on the relative velocity, $\overrightarrow{v}_{ij}=\overrightarrow{v}_{i}-%
\overrightarrow{v}_{j}$, and position $\overrightarrow{q}_{ij}$ (e.g.,
through the combination $\left( \overrightarrow{v}_{ij}\cdot \overrightarrow{%
q}_{ij}\right) ^{2}$).

\subsection{Collision\ Rules}

The coupling between chemistry and hydrodynamics can be captured simply by
considering atoms that carry a label (e.g., color) that can change during
collisions. Allowing for the non-conservation of energy gives a relatively
broad model that includes endo- and exothermic reactions. However, in the
interests of generality, the problem of modeling reactions that not only
violate energy conservation, but that also allow for the exchange, or even
loss, of mass will be considered.

The modelling of the collision rules in the case that mass is either
exchanged or lost upon collision is somewhat problematic. To understand why,
consider the usual arguments leading to specular collision rules in the case
that mass is invariant. Defining the total and relative momenta respectively
as%
\begin{eqnarray}
\overrightarrow{P} &=&\overrightarrow{p}_{1}+\overrightarrow{p}_{2} \\
\overrightarrow{p} &=&\overrightarrow{p}_{1}-\overrightarrow{p}_{2},  \notag
\end{eqnarray}%
the conservation of total momentum means that%
\begin{eqnarray}
\overrightarrow{P}^{\prime } &=&\overrightarrow{P}  \label{rule} \\
\overrightarrow{p}^{\prime } &=&\overrightarrow{p}+\overrightarrow{\gamma }%
_{l_{1}l_{2}}^{l_{1}^{\prime }l_{2}^{\prime }}  \notag
\end{eqnarray}%
where $\overrightarrow{\gamma }_{l_{1}l_{2}}^{l_{1}^{\prime }l_{2}^{\prime
}} $ $(x_{1,}x_{2})=-\overrightarrow{\gamma }_{l_{2}l_{1}}^{l_{2}^{\prime
}l_{1}^{\prime }}\left( x_{2},x_{1}\right) $ is to be determined. Second,
the energy balance equation can be written as%
\begin{equation}
E(x_{1}^{\prime },x_{2}^{\prime })+\delta E_{l_{1}l_{2}}^{l_{1}^{\prime
}l_{2}^{\prime }}(x_{1},x_{2})=E(x_{1}^{\prime },x_{2}^{\prime })
\end{equation}%
where $\delta E_{l_{1}l_{2}}^{l_{1}^{\prime }l_{2}^{\prime }}$ $%
(x_{1},x_{2}) $ is the energy lost during the collision. Substitution of Eq.(%
\ref{rule}) gives%
\begin{equation}
\left( \gamma _{l_{1}l_{2}}^{l_{1}^{\prime }l_{2}^{\prime }}\right)
^{2}+2\left( \overrightarrow{p}+\frac{m_{l_{2}}-m_{l_{1}}}{%
m_{l_{1}}+m_{l_{2}}}\overrightarrow{P}\right) \cdot \overrightarrow{\gamma }%
_{l_{1}l_{2}}^{l_{1}^{\prime }l_{2}^{\prime }}+8\mu _{l_{1}l_{2}}\delta
E_{l_{1}l_{2}}^{l_{1}^{\prime }l_{2}^{\prime }}=0  \label{coll1}
\end{equation}%
where the reduced mass is $\mu _{l_{1}l_{2}}=\frac{m_{l_{1}}m_{l_{2}}}{%
m_{l_{1}}+m_{l_{2}}}.$In $D$-dimensions, this gives one constraint on the $D$
independent components of $\overrightarrow{\gamma }_{l_{1}l_{2}}^{l_{1}^{%
\prime }l_{2}^{\prime }}$ so, for example, it fixes the magnitude of $%
\overrightarrow{\gamma }_{l_{1}l_{2}}^{l_{1}^{\prime }l_{2}^{\prime }}$ if
its direction is known: In one dimension, the problem is therefore solved.
In higher dimensions, the conservation of angular momentum gives the needed
additional constraint. This reads%
\begin{equation}
\overrightarrow{P}^{\prime }\times \frac{\overrightarrow{q}_{1}+%
\overrightarrow{q}_{2}}{2}+\frac{1}{2}\overrightarrow{p}_{12}^{\prime
}\times \overrightarrow{q}_{12}=\overrightarrow{P}\times \frac{%
\overrightarrow{q}_{1}+\overrightarrow{q}_{2}}{2}+\frac{1}{2}\overrightarrow{%
p}_{12}\times \overrightarrow{q}_{12}
\end{equation}%
or, using the conservation of total momentum,%
\begin{equation}
\overrightarrow{\gamma }_{l_{1}l_{2}}^{l_{1}^{\prime }l_{2}^{\prime }}\times 
\overrightarrow{q}_{12}=0
\end{equation}%
thus fixing the direction of $\overrightarrow{\gamma }_{l_{1}l_{2}}^{l_{1}^{%
\prime }l_{2}^{\prime }}$ as being along the line joining the centers of the
atoms. In this case, eq.(\ref{coll1}) gives%
\begin{equation}
\overrightarrow{\gamma }_{l_{1}l_{2}}^{l_{1}^{\prime }l_{2}^{\prime }}=2\mu
_{l_{1}l_{2}}\left( -\overrightarrow{v}_{12}\cdot \overrightarrow{q}_{12}-%
\sqrt{\left( \overrightarrow{v}_{12}\cdot \overrightarrow{q}_{12}\right)
^{2}-\frac{2}{\mu _{l_{1}l_{2}}}\delta E_{l_{1}l_{2}}^{l_{1}^{\prime
}l_{2}^{\prime }}}\right) \widehat{q}_{12}.
\end{equation}%
Taking $\delta E_{l_{1}l_{2}}^{l_{1}^{\prime }l_{2}^{\prime }}=0$ gives the
usual result for elastic hard spheres whereas setting the loss to be a fixed
fraction of the contribution to the rest-frame kinetic energy due to the
velocity along the line joining the atoms, $\delta
E_{l_{1}l_{2}}^{l_{1}^{\prime }l_{2}^{\prime }}=\lambda
_{l_{1}l_{2}}^{l_{1}^{\prime }l_{2}^{\prime }}\frac{1}{2}\mu
_{l_{1}l_{2}}\left( \overrightarrow{v}_{12}\cdot \overrightarrow{q}%
_{12}\right) ^{2}$, is the model used for inelastic hard spheres (i.e.,
granular fluids) and gives a coefficient of restitution $\alpha
_{l_{1}l_{2}}^{l_{1}^{\prime }l_{2}^{\prime }}=\sqrt{1-\lambda
_{l_{1}l_{2}}^{l_{1}^{\prime }l_{2}^{\prime }}}$.

When mass can be exchanged upon collision, it is useful to introduce the
total mass $M_{l_{1}l_{2}}=m_{l_{1}}+m_{l_{2}}$, the center of mass $%
\overrightarrow{Q}=\frac{m_{l_{1}}\overrightarrow{q}_{1}+m_{l_{2}}%
\overrightarrow{q}_{2}}{M_{l_{1}l_{2}}}$ and the center of mass velocity $%
\overrightarrow{V}=\overrightarrow{P}/M_{l_{1}l_{2}}$. Notice that even with
the conservation of total mass, the center of mass is not generally
invariant if mass is exchanged and the positions are kept fixed. So, the
conservation of angular momentum gives%
\begin{equation}
M_{l_{1}^{\prime }l_{2}^{\prime }}\overrightarrow{V}^{\prime }\times 
\overrightarrow{Q}^{\prime }+\mu _{l_{1}^{\prime }l_{2}^{\prime }}v^{\prime
}\times \overrightarrow{q}_{12}=M_{l_{1}l_{2}}\overrightarrow{V}\times 
\overrightarrow{Q}+\mu _{l_{1}l_{2}}\overrightarrow{v}\times \overrightarrow{%
q}_{12}
\end{equation}%
or, using the conservation of total momentum,%
\begin{equation}
M_{l_{1}l_{2}}\overrightarrow{V}\times \left( \overrightarrow{Q}^{\prime }-%
\overrightarrow{Q}\right) +\left( \mu _{l_{1}^{\prime }l_{2}^{\prime }}%
\overrightarrow{v}^{\prime }-\mu _{l_{1}l_{2}}\overrightarrow{v}\right)
\times \overrightarrow{q}_{12}=0.  \label{angular}
\end{equation}%
In general, this equation cannot be satisfied since it implies%
\begin{equation}
\left( \overrightarrow{V}\times \left( \overrightarrow{Q}^{\prime }-%
\overrightarrow{Q}\right) \right) \cdot \overrightarrow{q}_{12}=0
\end{equation}%
which is not generally true. The conclusion is that any collision rule which
does not leave invariant the center of mass will necessarily result in a
violation of the conservation of angular momentum at the microscopic level.
It is not possible to compensate by allowing the positions of the atoms to
shift during the collision since this could lead to overlapping
configurations involving a third atom. In fact, one would expect that the
inclusion of internal degrees of freedom, in particular of rotation of the
spherical atoms, would allow for a galilean-invariant collision law and this
will be explored at a later date. For present purposes, given that the
collision rule cannot be uniquely fixed by appealing to general principles,
the only recourse is to try to construct reasonable models. One possibility
is to conserve the angular momentum in the center of mass (CM)\ rest frame
since then $\overrightarrow{V}=0$ and angular momentum can indeed be
conserved. Another is to work in analogy to the case of invariant masses and
to require that either all momentum transfer be along $\overrightarrow{q}%
_{12}$ (so $\overrightarrow{p}^{\prime }=\overrightarrow{p}+2\mu
_{l_{1}l_{2}}\gamma _{l_{1}l_{2}}^{l_{1}^{\prime }l_{2}^{\prime }}\widehat{q}%
_{12}$) However, since the former is not Galilean invariant when mass is
transferred (since in a frame moving at velocity $\overrightarrow{u}$ the
relative momentum is $\overrightarrow{p}_{boosted}=\overrightarrow{p}+\left(
m_{l_{1}}-m_{l_{2}}\right) \overrightarrow{u}$), it is not clear how to
uniquely apply it. In fact, if one tries to enforce this constraint in the
CM frame, it gives the same result as fixing the angular momentum in the CM
frame. A second possibility is to demand that all velocity change be along $%
\overrightarrow{q}_{12}$ (so $\overrightarrow{v}^{\prime }=\overrightarrow{v}%
+\overline{\gamma }_{l_{1}l_{2}}^{l_{1}^{\prime }l_{2}^{\prime }}\widehat{q}%
_{12}$). For illustrative purposes, both options will be considered.

For the sake of generality, it is also interesting to consider the
consequences when mass is not only exchanged, but is lost. It is clear that
the previous considerations concerning the specification of the collision
rule under mass exchange apply to this case as well so that, again, a model
must be introduced in order to specify the relation between the lost mass
and the total momentum and energy. Without considering specific
applications, it is not clear that any unique conclusions can be drawn, so
by way of illustration, I will assume that the mass is carried away in such
a way that the total momentum in the CM frame is conserved. This means in
general that the law of conservation of momentum becomes%
\begin{equation}
\overrightarrow{P}^{\prime }=\overrightarrow{P}-\delta
m_{l_{1}l_{2}}^{l_{1}^{\prime }l_{2}^{\prime }}\overrightarrow{V}
\end{equation}%
where $\delta m_{l_{1}l_{2}}^{l_{1}^{\prime }l_{2}^{\prime
}}=m_{l_{1}}+m_{l_{2}}-m_{l_{1}^{\prime }}-m_{l_{2}^{\prime }}$. The model
adopted here is that the mass is carried away by $n_{l_{1}l_{2}}^{l_{1}^{%
\prime }l_{2}^{\prime }}$ particles with masses $m_{i}^{0}$ so that $\delta
m_{l_{1}l_{2}}^{l_{1}^{\prime }l_{2}^{\prime
}}=\sum_{i=1}^{n_{l_{1}l_{2}}^{l_{1}^{\prime }l_{2}^{\prime }}}m_{i}^{0}$
and with rest-frame velocities $\overrightarrow{v}_{i}^{0}$ satisfying $%
\sum_{i=1}^{n_{l_{1}l_{2}}^{l_{1}^{\prime }l_{2}^{\prime }}}\overrightarrow{v%
}_{i}^{0}=0$. For the case in which the angular momentum in the rest CM
frame is held constant, it is natural to also require that these particles
carry no net angular momentum.

Finally, some model must be specified for the energy lost (or gained), $%
\delta E_{l_{1}l_{2}}^{l_{1}^{\prime }l_{2}^{\prime }}$, which might include
contributions due to kinetic energy that is carried away by the lost mass
and energy lost (or gained) through other mechanisms (excitation of internal
degrees of freedom, radiation, exothermic and endothermic chemical
reactions...). Suppressing the species indices for a moment, the energy
differential can be written as the sum of two contributions, $\delta
E=\delta E_{m}+\delta E_{0}$, where the first is the energy carried away
with the lost mass and the second is due to any other inelastic processes.
In the CM\ rest frame, $\delta E_{m}=\sum_{i=1}^{n}\frac{1}{2}m_{i}\left( 
\overrightarrow{v}_{i}^{0}\right) ^{2}\equiv \overline{\delta E_{m}}$ so in
the lab frame, $\delta E_{m}=\sum_{i=1}^{n}\frac{1}{2}m_{i}\left( 
\overrightarrow{v}_{i}^{0}+\overrightarrow{V}\right) ^{2}=\overline{\delta
E_{m}}+\frac{1}{2}\delta mV^{2}$. Further, I assume, as is commonly done,
that the remaining energy loss (or gain) is frame independent (which means
in particular that it can only be a function of the relative velocity of the
colliding atoms). The energy balance equation therefore reads%
\begin{equation}
\frac{1}{2m_{l_{1}^{\prime }}}p_{1}^{\prime 2}+\frac{1}{2m_{l_{2}^{\prime }}}%
p_{2}^{\prime 2}+\overline{\delta E_{l_{1}l_{2}}^{l_{1}^{\prime
}l_{2}^{\prime }}}+\frac{1}{2}\delta m_{l_{1}l_{2}}^{l_{1}^{\prime
}l_{2}^{\prime }}V^{2}=\frac{1}{2m_{l_{1}}}p_{1}^{2}+\frac{1}{2m_{l_{2}}}%
p_{2}^{2}  \label{Ebal}
\end{equation}%
where it is understood that $\overline{\delta E_{l_{1}l_{2}}^{l_{1}^{\prime
}l_{2}^{\prime }}}=\overline{\delta E_{l_{1}l_{2}}^{l_{1}^{\prime
}l_{2}^{\prime }}}(x_{1},x_{2})$ is a galilean-invariant function of the
phases. This expression depends on the model for the lost mass (if any) but
is independent of any other assumptions concerning the collision rule.

From Eq.(\ref{angular}), conservation of angular momentum in the rest frame
then gives%
\begin{equation}
\overrightarrow{v}_{12}^{\prime }=\frac{\mu _{l_{1}l_{2}}}{\mu
_{l_{1}^{\prime }l_{2}^{\prime }}}\left( \overrightarrow{v}_{12}+\lambda
_{l_{1}l_{2}}^{l_{1}^{\prime }l_{2}^{\prime }}\widehat{q}\right)
\end{equation}%
for some scalar $\lambda _{l_{1}l_{2}}^{l_{1}^{\prime }l_{2}^{\prime
}}(x_{1},x_{2})$. Substituting into Eq.(\ref{Ebal}) gives

\begin{equation}
\lambda _{l_{1}l_{2}}^{l_{1}^{\prime }l_{2}^{\prime }}=-\overrightarrow{v}%
\cdot \widehat{q}+\sqrt{\left( \overrightarrow{v}\cdot \widehat{q}\right)
^{2}-\left( 1-\mu _{l_{1}^{\prime }l_{2}^{\prime }}/\mu _{l_{1}l_{2}}\right)
v^{2}-\frac{2\mu _{l_{1}^{\prime }l_{2}^{\prime }}}{\mu _{l_{1}l_{2}}^{2}}%
\overline{\delta E_{l_{1}l_{2}}^{l_{1}^{\prime }l_{2}^{\prime }}}-\frac{\mu
_{l_{1}^{\prime }l_{2}^{\prime }}}{\mu _{l_{1}l_{2}}^{2}}\delta
m_{l_{1}l_{2}}^{l_{1}^{\prime }l_{2}^{\prime }}V^{2}}
\end{equation}%
and one then finds that%
\begin{equation}
\overrightarrow{\gamma }_{l_{1}l_{2}}^{l_{1}^{\prime }l_{2}^{\prime }}=2\mu
_{l_{1}l_{2}}\lambda _{l_{1}l_{2}}^{l_{1}^{\prime }l_{2}^{\prime }}\widehat{q%
}_{12}+\left( \frac{m_{l_{1}^{\prime }}-m_{l_{1}}}{m_{l_{1}}+m_{l_{2}}}-%
\frac{m_{l_{2}^{\prime }}-m_{l_{2}}}{m_{l_{1}}+m_{l_{2}}}\right) 
\overrightarrow{P}.
\end{equation}%
Demanding that the change in the relative velocity be along the line joining
the atoms gives a very similar result%
\begin{equation}
\overrightarrow{v}_{12}^{\prime }=\overrightarrow{v}_{12}+\lambda
_{l_{1}l_{2}}^{l_{1}^{\prime }l_{2}^{\prime }}\widehat{q}_{12}
\end{equation}%
with%
\begin{equation}
\lambda _{l_{1}l_{2}}^{l_{1}^{\prime }l_{2}^{\prime }}=-\overrightarrow{v}%
_{12}\cdot \widehat{q}_{12}+\sqrt{\left( \overrightarrow{v}_{12}\cdot 
\widehat{q}_{12}\right) ^{2}-\left( 1-\mu _{l_{1}l_{2}}/\mu _{l_{1}^{\prime
}l_{2}^{\prime }}\right) v_{12}^{2}-\frac{2}{\mu _{l_{1}^{\prime
}l_{2}^{\prime }}}\overline{\delta E_{l_{1}l_{2}}^{l_{1}^{\prime
}l_{2}^{\prime }}}-\frac{\delta m_{l_{1}l_{2}}^{l_{1}^{\prime }l_{2}^{\prime
}}}{\mu _{l_{1}^{\prime }l_{2}^{\prime }}}V^{2}}
\end{equation}%
and%
\begin{equation}
\overrightarrow{\gamma }_{l_{1}l_{2}}^{l_{1}^{\prime }l_{2}^{\prime
}}=\left( \frac{\mu _{l_{1}^{\prime }l_{2}^{\prime }}}{\mu _{l_{1}l_{2}}}%
-1\right) \overrightarrow{p}_{12}+2\mu _{l_{1}^{\prime }l_{2}^{\prime
}}\lambda _{l_{1}l_{2}}^{l_{1}^{\prime }l_{2}^{\prime }}\widehat{q}%
_{12}+\left( \frac{m_{l_{1}^{\prime }}-m_{l_{2}^{\prime }}}{%
m_{l_{1}}+m_{l_{2}}}-\frac{\mu _{l_{1}^{\prime }l_{2}^{\prime }}}{\mu
_{l_{1}l_{2}}}\left( \frac{m_{l_{1}}-m_{l_{2}}}{m_{l_{1}}+m_{l_{2}}}\right)
\right) \overrightarrow{P}
\end{equation}%
Note that these two models coincide in the special case that the reduced
mass is invariant which obtains in one of two circumstances: the atomic
masses are invariant or if the atoms just exchange masses so that $%
m_{l_{1}^{\prime }}=m_{l_{2}}$ and vice-versa. In both cases, mass is
necessarily conserved, $\delta m_{l_{1}l_{2}}^{l_{1}^{\prime }l_{2}^{\prime
}}=0$, and all other conclusions are model-independent consequences of
galilean-invariance.

\bigskip

\subsection{The evolution of phase functions}

The dynamics of any hard-sphere model consists of free streaming interrupted
by binary collisions. In non-reacting fluids, the collisions lead to an
instantaneous change of the velocities of the colliding atoms. The
generalization to the reacting fluid only requires that the chemical species
labels, and hence the masses and any other species-specific properties, to
be viewed as dynamical variables as well and, so, as part of an enlarged
phase space.

Two atoms, say atoms $i$ and $j$, collide at time $\tau _{ij}$ when their
centers are separated by their relative hard-sphere diameter $\sigma
_{l_{i}l_{j}}$%
\begin{equation}
\left| \overrightarrow{q}_{i}(\tau _{ij})-\overrightarrow{q}_{j}(\tau
_{ij})\right| =q_{ij}^{2}(\tau _{ij})=\sigma _{l_{i}l_{j}}^{2}  \label{cq}
\end{equation}%
where, e.g., for additive models, the relative hard sphere diameter is
simply the sum of the atoms' radii $\sigma _{l_{i}l_{j}}=\frac{1}{2}\sigma
_{l_{j}}+\frac{1}{2}\sigma _{l_{j}}$. The atoms do not have to all have the
same size (e.g., an acceptable possibility is that different species have
different sizes but chemical reactions always transform atoms from a species
of a given size to other species of the same size). An exceptional
possibility in which size could change is one in which atoms only get
smaller upon collision: this might be useful to model certain granular
materials that fragment upon collision (e.g., the ice composing the rings of
Saturn) and could be handled within the present formalism as long as the
position of the center of mass of each atom is invariant. From Eq.(\ref{cq}%
), one has that the time of collision is 
\begin{equation}
\tau _{ij}\left( \Gamma \right) =-\frac{1}{v_{ij}^{2}}\overrightarrow{v}%
_{ij}\cdot \overrightarrow{q}_{ij}-\frac{1}{v_{ij}^{2}}\sqrt{\left( 
\overrightarrow{v}_{ij}\cdot \overrightarrow{q}_{ij}\right)
^{2}-v_{ij}^{2}\left( q_{ij}^{2}-\sigma _{l_{i}l_{j}}^{2}\right) }
\end{equation}%
where the sign has been chosen according to give the physical solution. If
the right hand side is imaginary, then no collision takes place for the
given initial conditions. This aspect of the dynamics is independent of what
actually happens after the collision and is the reason that the structure of
the pseudo-Liouville equation is independent of the collision rule. The
pseudo-Liouville equation describing the time evolution of an arbitrary
phase function, $A\left( \Gamma ;t\right) =A\left(
x_{1},l_{1},...x_{N},l_{N};t\right) $ then follows immediately by analogy
with the non-reacting fluid and is 
\begin{eqnarray}
\frac{d}{dt}A &=&\frac{\partial }{\partial t}A+\widehat{\mathcal{L}}_{+}A \\
\widehat{\mathcal{L}}_{+} &=&\sum_{i}\overset{\cdot }{x}_{i}\frac{\partial }{%
\partial x_{i}}+\sum_{i<j}\widehat{\mathcal{T}}_{+}\left( ij\right)  \notag
\end{eqnarray}%
where the binary collision operators are 
\begin{equation}
\widehat{\mathcal{T}}_{+}\left( ij\right) =-\overrightarrow{q}_{ij}\cdot 
\overrightarrow{v}_{ij}\delta \left( q_{ij}-\sigma _{l_{i}l_{j}}\right)
\Theta \left( -\widehat{q}_{ij}\cdot \overrightarrow{v}_{ij}\right) \left(
\sum_{l_{i}^{\prime }l_{j}^{\prime }}\mathcal{M}_{l_{i}l_{j}}^{l_{i}^{\prime
}l_{j}^{\prime }}\left( x_{ij}\right) \widehat{b}_{l_{i}l_{j}}^{l_{i}^{%
\prime }l_{j}^{\prime }}-1\right) .
\end{equation}%
As discussed in Appendix \ref{AppT}, this can be derived directly for a
system of two atoms by writing the exact solution to the two-body problem
and differentiating; the generalization to $N$-atoms follows immediately due
to the fact that only binary collisions occur. Here, $\mathcal{M}%
_{l_{i}l_{j}}^{l_{i}^{\prime }l_{j}^{\prime }}\left( x_{i}{}_{j}\right) $ is
a random matrix which, in any realization, takes on the value $1$ for some
single combination of $l_{i}^{\prime },l_{j}^{\prime }$ and is zero
otherwise and which is distributed according to 
\begin{equation}
\left\langle \mathcal{M}_{l_{i}l_{j}}^{l_{i}^{\prime }l_{j}^{\prime }}\left(
x_{ij}\right) \right\rangle _{react}=K_{l_{i}l_{j}}^{l_{i}^{\prime
}l_{j}^{\prime }}\left( x_{ij}\right) .  \label{stoch}
\end{equation}%
(The notation used here indicates stochastic quantities by means of
Calligraphic type and uses carets to denote operators and averages over the
stochastic process are denoted as $\left\langle ...\right\rangle _{react}$.)
For a non-reacting system, it becomes $\mathcal{M}_{l_{i}l_{j}}^{l_{i}^{%
\prime }l_{j}^{\prime }}\left( x_{ij}\right) =\delta _{l_{i}^{\prime
}l_{i}}\delta _{l_{j}^{\prime }l_{j}}$. The only other formal difference
from the non-reacting case is that the momentum transfer operator, $\widehat{%
b}_{l_{i}l_{j}}^{l_{i}^{\prime }l_{j}^{\prime }}$, has the effect of
altering both the mechanical variables and the species labels. So, just as
this operator instantaneously changes the position in phase space of the $i$%
th atom from $x_{i}(t_{-})$ before a collision at time $t$ to $%
x_{i}(t_{+})=x_{i}^{\prime }(t_{-})$ it also instantaneously alters the
species labels from $l_{i}\left( t_{-}\right) $ to $l_{i}\left( t_{+}\right)
=l_{i}^{\prime }\left( t_{-}\right) $ the difference being that $%
x_{i}^{\prime }(t_{-})$ is a deterministic function of $x_{i}(t_{-})$ and $%
x_{j}(t_{-})$ whereas the evolution of $l_{i}^{\prime }(t_{-})$ is
stochastic. For phase functions which have no explicit time dependence, the
Liouville equation can be formally solved to get%
\begin{equation}
\mathcal{A}(\Gamma ,t)=\exp \left( \widehat{\mathcal{L}}_{+}t\right)
A(\Gamma )
\end{equation}%
which has the meaning that the system evolves from the initial phase $\Gamma 
$.

The most important difference from the non-reacting system appears in the
evaluation of statistical averages. In the presence of reactions there are
two statistical processes that must be considered: the distribution of
initial conditions and the stochastic process that alters species labels at
the collisions. For a given distribution of initial conditions $\rho
^{(0)}\left( \Gamma \right) =\rho _{l_{1}l_{2}...l_{N}}^{(0)}\left(
x_{1},x_{2},...x_{N}\right) $ (giving the probability that the first atom
begins with species $l_{1}$ and phase $x_{1}$, and so), one has%
\begin{equation}
\left\langle A;t\right\rangle =\int d\Gamma \;\rho ^{(0)}\left( \Gamma
\right) \left\langle \mathcal{A}(t)\right\rangle _{react}=\int d\Gamma
\;\rho ^{(0)}\left( \Gamma \right) \left\langle \exp \left( \widehat{%
\mathcal{L}}_{+}t\right) A(\Gamma )\right\rangle _{react}
\end{equation}%
and the notation should be understood as implying a sum over the initial
species labels%
\begin{eqnarray}
&&\int d\Gamma \;\rho ^{(0)}\left( \Gamma \right) \left\langle \exp \left( 
\widehat{\mathcal{L}}_{+}t\right) A(\Gamma )\right\rangle _{react} \\
&\equiv &\sum_{l_{1}...l_{N}}\int dx_{1}...dx_{N}\;\rho
_{l_{1}...l_{N}}^{(0)}\left( x_{1},...x_{N}\right) \left\langle \exp \left( 
\widehat{\mathcal{L}}_{+}t\right) \right\rangle _{react}A(\Gamma )  \notag
\end{eqnarray}%
where $A(\Gamma )$ can be taken outside of the average over reactions since
it depends only on the initial conditions. Now, since each collision
involves an independent stochastic process, it follows that $\left\langle
\exp \left( \widehat{\mathcal{L}}_{+}t\right) \right\rangle _{react}=\exp
\left( \left\langle \widehat{\mathcal{L}}_{+}\right\rangle _{react}t\right) $
which is evaluated using Eq.(\ref{stoch}). Thus, the time averages become%
\begin{equation}
\left\langle A;t\right\rangle =\sum_{l_{1}...l_{N}}\int
dx_{1}...dx_{N}\;\rho _{l_{1}...l_{N}}^{(0)}\left( x_{1},...x_{N}\right)
\exp \left( \widehat{L}_{+}t\right) A(\Gamma )  \label{av}
\end{equation}%
with the deterministic operator%
\begin{equation}
\widehat{L}_{+}=\sum_{i}\overset{\cdot }{x}_{i}\frac{\partial }{\partial
x_{i}}+\sum_{i<j}\widehat{T}_{+}\left( ij\right) 
\end{equation}%
and the (reaction-averaged) collision operators are%
\begin{equation}
\widehat{T}_{+}\left( ij\right) =-\overrightarrow{q}_{ij}\cdot 
\overrightarrow{v}_{ij}\delta \left( q_{ij}-\sigma _{l_{i}l_{j}}\right)
\Theta \left( -\widehat{q}_{ij}\cdot \overrightarrow{v}_{ij}\right) \left(
\sum_{l_{i}^{\prime }l_{j}^{\prime }}K_{l_{i}l_{j}}^{l_{i}^{\prime
}l_{j}^{\prime }}\left( x_{ij}\right) \widehat{b}_{l_{i}l_{j}}^{l_{i}^{%
\prime }l_{j}^{\prime }}-1\right) .
\end{equation}%
This shows that, from the point of view of evaluating the statistical
averages, it suffices to work with the deterministic dynamics defined by $%
\widehat{L}_{+}$ which no longer treats the species labels as discrete
stochastic variables. Instead, the phase functions are at all times averaged
over the reactions and so do not explicitly represent dynamical quantities
as might be realized in a computer simulation. In fact, they correspond to
the average result of an ensemble of simulations, all beginning with
identical initial conditions, but differing in the realization of the
reaction process $\mathcal{M}_{l_{i}l_{j}}^{l_{i}^{\prime }l_{j}^{\prime
}}\left( x_{i}{}_{j}\right) $.

\subsection{The evolution of the distribution function}

The adjoint $\widehat{L}_{+}^{A}$ of the Liouville operator $\widehat{L}_{+}$
is defined as%
\begin{equation}
\int d\Gamma \;B\left( \Gamma \right) \widehat{L}_{+}A\left( \Gamma \right)
=\int d\Gamma \;\left( \widehat{L}_{+}^{A}B\left( \Gamma \right) \right)
A\left( \Gamma \right)  \label{adjoint_def}
\end{equation}%
from which one finds (see appendix \ref{AppA})%
\begin{equation}
\widehat{L}_{+}^{A}=-\sum_{i}\overset{\cdot }{x}_{i}\frac{\partial }{%
\partial x_{i}}+\sum_{i<j}\widehat{T}_{+}^{A}\left( ij\right)
\end{equation}%
with the adjoint collision operator%
\begin{equation}
\widehat{T}_{+}^{A}\left( ij\right) =-\left[ \sum_{a,b}J_{ab}^{l_{i}l_{j}}%
\left( x_{i},x_{j}\right) \left( \widehat{b}_{ab}^{l_{i}l_{j}}\right)
^{-1}K_{ab}^{l_{i}l_{j}}\left( x_{ij}\right) -1\right] \Theta \left( -%
\overrightarrow{v}_{ij}\cdot \overrightarrow{q}_{ij}\right) \delta \left(
q_{ij}-\sigma _{l_{i}l_{j}}\right) \overrightarrow{v}_{ij}\cdot 
\overrightarrow{q}_{ij}
\end{equation}%
with%
\begin{equation}
J_{ab}^{l_{1}l_{2}}\left( x_{i},x_{j}\right) =\left| \frac{\partial \left(
\left( \widehat{b}_{ab}^{l_{i}l_{j}}\right) ^{-1}x_{i},\left( \widehat{b}%
_{ab}^{l_{i}l_{j}}\right) ^{-1}x_{j}\right) }{\partial \left(
x_{i},x_{j}\right) }\right| ^{-1}.
\end{equation}%
Here, the operator $\left( \widehat{b}_{ab}^{l_{i}l_{j}}\right) ^{-1}$ is
the inverse of $\widehat{b}_{ab}^{l_{i}l_{j}}$ both in terms of the change
of the mechanical variables as well as the species labels so that for an
arbitrary function $\left( \widehat{b}_{ab}^{l_{i}l_{j}}\right) ^{-1}B\left(
x_{i},l_{i};x_{j},l_{j}\right) =B\left( \left( \widehat{b}%
_{ab}^{l_{i}l_{j}}\right) ^{-1}x_{i},a;\left( \widehat{b}_{ab}^{l_{i}l_{j}}%
\right) ^{-1}x_{j},b\right) $. To illustrate the structure of this operator,
consider the case of inelastic hard spheres used to model granular fluids.
Specializing to a single species, one has%
\begin{equation}
\overrightarrow{v}_{ij}^{\prime }=\widehat{b}\overrightarrow{v}_{ij}=%
\overrightarrow{v}_{ij}-\left( 1+\alpha \right) \overrightarrow{v}_{ij}\cdot 
\widehat{q}_{ij},
\end{equation}%
where $\alpha $ is a constant, from which it follows that 
\begin{equation}
\overrightarrow{v}_{ij}=\widehat{b}^{-1}\overrightarrow{v}_{ij}^{\prime }=%
\overrightarrow{v}_{ij}^{\prime }-\left( \frac{1+\alpha }{\alpha }\right) 
\overrightarrow{v}_{ij}^{\prime }\cdot \widehat{q}_{ij}
\end{equation}%
giving%
\begin{equation}
\left| \frac{\partial \left( \widehat{b}^{-1}x_{i},\widehat{b}%
^{-1}x_{j}\right) }{\partial \left( x_{i},x_{j}\right) }\right| =\left|
1-\left( \frac{1+\alpha }{\alpha }\right) \right| =\frac{1}{\alpha }
\end{equation}%
so that 
\begin{equation}
\widehat{T}_{+}^{A}\left( ij\right) B\left( x_{i},l_{i};x_{j},l_{j}\right) =-%
\left[ \frac{1}{\alpha }\widehat{b}^{-1}-1\right] \Theta \left( -%
\overrightarrow{v}_{ij}\cdot \overrightarrow{q}_{ij}\right) \delta \left(
q_{ij}-\sigma _{l_{i}l_{j}}\right) \overrightarrow{v}_{ij}\cdot 
\overrightarrow{q}_{ij}B\left( x_{i},l_{i};x_{j},l_{j}\right) .
\end{equation}%
which is the usual result\cite{HCSLiouville}.

An important generalization of this result concerns the case that the
inverse transformation $\widehat{b}_{ab}^{l_{i}l_{j}}x_{i}$ is not unique.
This can happen even in the single species, inelastic case if the
coefficient of restitution depends on the velocities. For example, if $%
\alpha =\alpha \left( \overrightarrow{v}_{ij}\cdot \widehat{q}_{ij}\right) $
then the inverse collision rule is determined by solving%
\begin{equation}
\overrightarrow{v}_{ij}^{\prime }\cdot \widehat{q}_{ij}=-\alpha \left( 
\overrightarrow{v}_{ij}\cdot \widehat{q}_{ij}\right) \overrightarrow{v}%
_{ij}\cdot \widehat{q}_{ij}
\end{equation}%
which may or may not have a unique solution. In the latter case, $\widehat{T}%
_{+}^{A}$ must be written in terms of a sum over the various branches and
must include step functions which restrict the domain of integration in Eq.(%
\ref{adjoint_def}) to the appropriate domain for each branch. In practical
calculations, it is usually most convenient to recast integrals over $%
\widehat{T}_{+}^{A}\left( ij\right) $ into integrals involving $\widehat{T}%
_{+}\left( ij\right) $ so as to avoid this complication.

Given the adjoint operator $\widehat{L}_{+}^{A}$, Eq.(\ref{av}) can be
written as%
\begin{equation}
\left\langle A;t\right\rangle =\int d\Gamma \;\left( \exp \left(
L^{A}t\right) \rho ^{(0)}\left( \Gamma \right) \right) A\left( \Gamma
\right) \equiv \int d\Gamma \;\rho \left( \Gamma ;t\right) A\left( \Gamma
\right)
\end{equation}%
where the second equality defines the time-dependent distribution function.
It's time dependence is given by the pseudo-Liouville equation 
\begin{equation}
\left( \frac{\partial }{\partial t}+\sum_{i}\overset{\cdot }{x}_{i}\frac{%
\partial }{\partial x_{i}}+\sum_{i<j}\overline{T}_{-}\left( ij\right)
\right) \rho =0  \label{Liouville}
\end{equation}%
where in the standard notation\cite{McLennan}%
\begin{equation}
\overline{T}_{-}\left( ij\right) =-T_{+}^{A}\left( ij\right) .
\end{equation}

The Born-Bogoliubov-Green-Kirkwood-Yvon (BBGKY) hierarchy follows
immediately from the Liouville equation. Defining the reduced distribution
functions as 
\begin{equation}
f_{l_{1}...l_{m}}\left( x_{1}...x_{m}\right) =\frac{N!}{\left( N-m\right) !}%
\sum_{l_{m+1}...l_{N}}\int dx_{m+1}...dx_{N}\;\rho \left( \Gamma \right) .
\end{equation}%
and integrating the pseudo-Liouville equation over $x_{m+1}...x_{N}$ and
summing over the corresponding species labels gives the $m$-th equation of
the hierarchy%
\begin{eqnarray}
&&\left( \frac{\partial }{\partial t}+\sum_{i=1}^{m}\overrightarrow{v}%
_{i}\cdot \frac{\partial }{\partial \overrightarrow{q}_{i}}+\sum_{1\leq
i<j\leq m}\overline{T}_{-}\left( ij\right) \right) f_{l_{1}...l_{m}}\left(
x_{1}...x_{m}\right)  \label{BBGKY} \\
&=&-\sum_{i=1}^{m}\sum_{l_{m+1}}\int dx_{m+1}\overline{T}_{-}\left(
im+1\right) f_{l_{1}...l_{m+1}}\left( x_{1}...x_{m+1}\right) .  \notag
\end{eqnarray}%
The first equation of the hierarchy is the starting point for the Enskog
kinetic theory as described below.

\section{Exact balance equations}

Now consider the phenomenology of the reacting fluid which is expressed in
terms of the macroscopic hydrodynamic fields. The results presented here are
derived using only the general form of the collision rule, eq.(\ref{rule}),
and the microscopic energy balance equation, eq.(\ref{Ebal}) so that the
only assumptions made with respect to the collision model are those
concerning the energy transported by any lost mass.

The local fields of interest are the number fractions%
\begin{equation}
n_{l}\left( \overrightarrow{r},t\right) =\left\langle \sum_{i}\delta \left( 
\overrightarrow{r}-\overrightarrow{q}_{i}\right) \delta
_{ll_{i}};t\right\rangle =\int d\overrightarrow{v}_{1}\;f_{l}\left( 
\overrightarrow{r},\overrightarrow{v}_{1};t\right) ,
\end{equation}%
and the mass, momentum and energy densities, defined respectively as%
\begin{eqnarray}
\rho \left( \overrightarrow{r},t\right) &=&\sum_{l}m_{l}n_{l}\left( 
\overrightarrow{r},t\right) \\
\rho \left( \overrightarrow{r},t\right) \overrightarrow{u}(\overrightarrow{r}%
,t) &=&\left\langle \sum_{i}m_{l_{i}}\overrightarrow{v}_{i}\delta \left( 
\overrightarrow{r}-\overrightarrow{q}_{i}\right) ;t\right\rangle
=\sum_{l}m_{l}\int d\overrightarrow{v}_{1}\;\overrightarrow{v}%
_{1}f_{l}\left( \overrightarrow{r},\overrightarrow{v}_{1};t\right)  \notag \\
\frac{D}{2}n\left( \overrightarrow{r},t\right) k_{B}T\left( \overrightarrow{r%
},t\right) &=&\left\langle \sum_{i}\frac{1}{2}m_{l_{i}}V_{i}^{2}\delta
\left( \overrightarrow{r}-\overrightarrow{q}_{i}\right) ;t\right\rangle
=\sum_{l}\frac{1}{2}m_{l}\int d\overrightarrow{v}_{1}\;V_{1}^{2}f_{l}\left( 
\overrightarrow{r},\overrightarrow{v}_{1};t\right)  \notag
\end{eqnarray}%
where the excess velocity is $\overrightarrow{V}_{i}(t)=\overrightarrow{v}%
_{i}(t)-\overrightarrow{u}\left( \overrightarrow{q}_{i},t\right) $and the
total number density is 
\begin{equation}
n\left( \overrightarrow{r},t\right) =\sum_{l}n_{l}\left( \overrightarrow{r}%
,t\right) .
\end{equation}%
It is also convenient to introduce the number fractions, or concentrations, $%
x_{l}\left( \overrightarrow{r},t\right) =n_{l}\left( \overrightarrow{r}%
,t\right) /n\left( \overrightarrow{r},t\right) $. The balance equations for
these quantities follow directly from their definitions and the first
equation of the BBGKY hierarchy. The details of the derivation are given in
appendix \ref{AppB} and only the results summarized here.

\subsection{Number. mass and concentration}

Integrating over the positions and velocities gives the balance equation for
the local partial number density%
\begin{equation}
\frac{\partial }{\partial t}n_{l}+\overrightarrow{\nabla }\cdot \left( 
\overrightarrow{u}n_{l}\right) +\overrightarrow{\nabla }\cdot 
\overrightarrow{j}_{l}=S_{l}^{(n)}  \label{n1}
\end{equation}%
with the source%
\begin{eqnarray}
S_{l}^{(n)}\left( \overrightarrow{r},t\right)  &=&-\frac{1}{2}%
\sum_{abl_{1}l_{2}}\int dx_{1}dx_{2}\;\left( \overrightarrow{q}_{12}\cdot 
\overrightarrow{v}_{12}\right) \delta \left( q_{12}-\sigma
_{l_{1}l_{2}}\right) \Theta \left( -\widehat{q}_{12}\cdot \overrightarrow{v}%
_{12}\right)   \label{n2} \\
&&\times f_{l_{1}l_{2}}\left( x_{1},x_{2};t\right) \delta \left( 
\overrightarrow{r}-\overrightarrow{q}_{1}\right) K_{l_{1}l_{2}}^{ab}\left(
x_{12}\right) \left( \delta _{al}+\delta _{bl}-\delta _{ll_{1}}-\delta
_{ll_{2}}\right)   \notag
\end{eqnarray}%
and the number current $\overrightarrow{j}_{l}=\overrightarrow{j}_{l}^{K}+%
\overrightarrow{j}_{l}^{V}$ with%
\begin{equation}
\overrightarrow{j}_{l}^{K}\left( \overrightarrow{r},t\right) =\int d%
\overrightarrow{v}_{1}\;f_{l}\left( \overrightarrow{r},\overrightarrow{v}%
_{1},t\right) \overrightarrow{V}_{1}  \label{n3}
\end{equation}%
and%
\begin{eqnarray}
\overrightarrow{j}_{l}^{V}\left( \overrightarrow{r},t\right)  &=&-\frac{1}{2}%
\sum_{abl_{1}l_{2}}\int dx_{1}dx_{2}\;\overrightarrow{q}_{12}\left( 
\overrightarrow{q}_{12}\cdot \overrightarrow{v}_{12}\right) \delta \left(
q_{12}-\sigma _{l_{1}l_{2}}\right) \Theta \left( -\widehat{q}_{12}\cdot 
\overrightarrow{v}_{12}\right)   \label{n4} \\
&&\times f_{l_{1}l_{2}}\left( x_{1},x_{2};t\right) K_{l_{1}l_{2}}^{ab}\left(
x_{12}\right) \left( \delta _{al}-\delta _{bl}-\delta _{ll_{1}}+\delta
_{ll_{2}}\right)   \notag \\
&&\times \int_{0}^{1}dx\;\delta \left( \overrightarrow{r}-x\overrightarrow{q}%
_{1}-\left( 1-x\right) \overrightarrow{q}_{2}\right) .  \notag
\end{eqnarray}%
The source term represents the gain or loss of atoms of type $l$ due to
chemical reactions. The kinetic part of the number current is familiar from
the study of multi-component, non-reacting systems\cite{McLennan} where it
takes the form $\overrightarrow{j}_{l}^{K}=\sum_{j}D_{lj}\overrightarrow{%
\nabla }n_{j}+L_{l}\overrightarrow{\nabla }T+o\left( \nabla ^{2}\right) $
and, e.g., gives rise to Fick's law when substituted into eq.(\ref{n1}) .
Here, it is seen that this diffusive current is enhanced by a second
contribution, eq.(\ref{n4}), that arises solely from the reactions (i.e., it
vanishes if $K_{l_{1}l_{2}}^{ab}=\delta _{al_{1}}\delta _{bl_{2}}$). This is
due to the transport of type-$l$ atoms due to the reaction process. The
conservation of total number density immediately follows by summing over the
species label%
\begin{equation}
\frac{\partial }{\partial t}n+\overrightarrow{\nabla }\cdot \left( 
\overrightarrow{u}n\right) +\overrightarrow{\nabla }\cdot \sum_{l}%
\overrightarrow{j}_{l}^{K}=0
\end{equation}%
where the sum over the species of the collisional contributions to the
number current vanishes. Similarly, multiplying by $m_{l}$ and then summing
gives the balance equation for local mass density%
\begin{equation}
\frac{\partial }{\partial t}\rho +\overrightarrow{\nabla }\cdot \left( 
\overrightarrow{u}\rho \right) +\overrightarrow{\nabla }\cdot 
\overrightarrow{Q}=S^{(\rho )},
\end{equation}%
where, the mass flux is 
\begin{eqnarray}
\overrightarrow{Q}\left( \overrightarrow{r},t\right)  &=&-\frac{1}{2}%
\sum_{abl_{1}l_{2}}\int dx_{1}dx_{2}\;\overrightarrow{q}_{12}\left( 
\overrightarrow{q}_{12}\cdot \overrightarrow{v}_{12}\right) \delta \left(
q_{12}-\sigma _{l_{1}l_{2}}\right) \Theta \left( -\widehat{q}_{12}\cdot 
\overrightarrow{v}_{12}\right)  \\
&&\times f_{l_{1}l_{2}}\left( x_{1},x_{2};t\right) K_{l_{1}l_{2}}^{ab}\left(
x_{12}\right) \left( m_{a}-m_{b}-m_{l_{1}}+m_{l_{2}}\right)   \notag \\
&&\times \int_{0}^{1}dx\;\delta \left( \overrightarrow{r}-x\overrightarrow{q}%
_{1}-\left( 1-x\right) \overrightarrow{q}_{2}\right) ,  \notag
\end{eqnarray}%
which vanishes if no mass is transported during collisions, and the mass
source term is%
\begin{eqnarray}
S_{l}^{(\rho )}\left( \overrightarrow{r},t\right)  &=&\frac{1}{2}%
\sum_{abl_{1}l_{2}}\int dx_{1}dx_{2}\;\left( \overrightarrow{q}_{12}\cdot 
\overrightarrow{v}_{12}\right) \delta \left( q_{12}-\sigma
_{l_{1}l_{2}}\right) \Theta \left( -\widehat{q}_{12}\cdot \overrightarrow{v}%
_{12}\right)  \\
&&\times f_{l_{1}l_{2}}\left( x_{1},x_{2};t\right) \delta \left( 
\overrightarrow{r}-\overrightarrow{q}_{1}\right) K_{l_{1}l_{2}}^{ab}\left(
x_{12}\right) \delta m_{l_{1}l_{2}}^{ab},  \notag
\end{eqnarray}%
which is only nonzero if the collisions do not conserve mass. Finally, using
the definition of the concentrations, $x_{l}=n_{l}/n$, the reaction equation
is found to be 
\begin{equation}
\frac{\partial }{\partial t}x_{l}+\overrightarrow{u}\cdot \overrightarrow{%
\nabla }x_{l}+n^{-1}\left[ \overrightarrow{\nabla }\cdot \overrightarrow{j}%
_{l}-x_{l}\overrightarrow{\nabla }\cdot \sum_{l}\overrightarrow{j}_{l}^{K}%
\right] =n^{-1}S_{l}^{(n)}
\end{equation}%
where the term on the right is now identified as the reaction rate.

\subsection{Momentum and velocity fields}

The balance equation for the local momentum, written in terms of the local
velocity, is%
\begin{equation}
\frac{\partial }{\partial t}\rho \overrightarrow{u}+\overrightarrow{\nabla }%
\cdot \left( \rho \overrightarrow{u}\overrightarrow{u}\right) +%
\overrightarrow{\nabla }\cdot \left( \overleftrightarrow{P}+\overrightarrow{Q%
}\overrightarrow{u}\right) =\overline{\overrightarrow{S}}^{(p)}+%
\overrightarrow{u}S_{l}^{(\rho )}
\end{equation}%
with the pressure tensor $\overleftrightarrow{P}=\overleftrightarrow{P}^{K}+%
\overleftrightarrow{P}^{V}+\overleftrightarrow{P}^{M}$ where the kinetic
contribution is 
\begin{equation}
\overleftrightarrow{P}^{K}\left( \overrightarrow{r},t\right)
=\sum_{l}m_{l}\int d\overrightarrow{v}_{1}\;f_{l}\left( \overrightarrow{r},%
\overrightarrow{v}_{1},t\right) \overrightarrow{V}_{1}\overrightarrow{V}_{1},
\end{equation}%
and the collisional contribution is%
\begin{eqnarray}
\overleftrightarrow{P}^{V}\left( \overrightarrow{r},t\right)  &=&-\frac{1}{2}%
\sum_{l_{1}l_{2}l_{1}^{\prime }l_{2}^{\prime }}\int dx_{1}dx_{2}\;%
\overrightarrow{q}_{12}\left( \overrightarrow{q}_{12}\cdot \overrightarrow{v}%
_{12}\right) \delta \left( q_{12}-\sigma _{l_{1}l_{2}}\right) \Theta \left( -%
\widehat{q}_{12}\cdot \overrightarrow{v}_{12}\right)  \\
&&\times f_{l_{1}l_{2}}\left( x_{1},x_{2};t\right)
K_{l_{1}l_{2}}^{l_{1}^{\prime }l_{2}^{\prime }}\left( x_{12}\right) 
\widetilde{\overrightarrow{\gamma }}_{l_{1}l_{2}}^{l_{1}^{\prime
}l_{2}^{\prime }}\int_{0}^{1}dx\;\delta \left( \overrightarrow{r}-x%
\overrightarrow{q}_{1}-\left( 1-x\right) \overrightarrow{q}_{2}\right)  
\notag
\end{eqnarray}%
where $\widetilde{\overrightarrow{\gamma }}_{l_{1}l_{2}}^{l_{1}^{\prime
}l_{2}^{\prime }}$ is the change of momentum in the rest frame%
\begin{equation}
\overrightarrow{\gamma }_{l_{1}l_{2}}^{l_{1}^{\prime }l_{2}^{\prime }}=%
\widetilde{\overrightarrow{\gamma }}_{l_{1}l_{2}}^{l_{1}^{\prime
}l_{2}^{\prime }}+\left( m_{l_{1}^{\prime }}-m_{l_{1}}-m_{l_{2}^{\prime
}}-m_{l_{2}}\right) \overrightarrow{V}_{12}.
\end{equation}%
Finally, the contribution from the instantaneous exchange of mass is%
\begin{eqnarray}
\overleftrightarrow{P}^{M} &=&-\frac{1}{2}\sum_{l_{1}l_{2}l_{1}^{\prime
}l_{2}^{\prime }}\int dx_{1}dx_{2}\;\overrightarrow{q}_{12}\left( 
\overrightarrow{q}_{12}\cdot \overrightarrow{v}_{12}\right) \delta \left(
q_{12}-\sigma _{l_{1}l_{2}}\right) \Theta \left( -\widehat{q}_{12}\cdot 
\overrightarrow{v}_{12}\right)  \\
&&\times f_{l_{1}l_{2}}\left( x_{1}x_{2}\right)
K_{l_{1}l_{2}}^{l_{1}^{\prime }l_{2}^{\prime }}\left( x_{12}\right) \left( 
\overrightarrow{V}_{12}-\overrightarrow{u}\right) \left( m_{l_{1}^{\prime
}}-m_{l_{1}}-m_{l_{2}^{\prime }}-m_{l_{2}}\right)   \notag \\
&&\times \int_{0}^{1}dx\;\delta \left( \overrightarrow{r}-x\overrightarrow{q}%
_{1}-\left( 1-x\right) \overrightarrow{q}_{2}\right)   \notag
\end{eqnarray}%
The source terms arise due to momentum being carried away by the lost mass
and the new term is given by%
\begin{eqnarray}
\overline{\overrightarrow{S}}^{(p)}\left( \overrightarrow{r},t\right)  &=&%
\frac{1}{2}\sum_{l_{1}l_{2}l_{1}^{\prime }l_{2}^{\prime }}\int
dx_{1}dx_{2}\;\left( \overrightarrow{q}_{12}\cdot \overrightarrow{v}%
_{12}\right) \delta \left( q_{12}-\sigma _{l_{1}l_{2}}\right) \Theta \left( -%
\widehat{q}_{12}\cdot \overrightarrow{v}_{12}\right)  \\
&&\times \left( \overrightarrow{V}_{12}-\overrightarrow{u}\right)
f_{l_{1}l_{2}}\left( x_{1},x_{2};t\right) K_{l_{1}l_{2}}^{l_{1}^{\prime
}l_{2}^{\prime }}\left( x_{12}\right) \delta m_{l_{1}l_{2}}^{l_{1}^{\prime
}l_{2}^{\prime }}\delta \left( \overrightarrow{r}-\overrightarrow{q}%
_{1}\right) .  \notag
\end{eqnarray}%
By using the balance equation for total mass density the equation of motion
for the velocity field is found to be 
\begin{equation}
\frac{\partial }{\partial t}\overrightarrow{u}+\overrightarrow{u}\cdot 
\overrightarrow{\nabla }\overrightarrow{u}+\rho ^{-1}\left( \overrightarrow{%
\nabla }\cdot \overleftrightarrow{P}+\overrightarrow{Q}\cdot \overrightarrow{%
\nabla }\overrightarrow{u}\right) =\rho ^{-1}\overline{\overrightarrow{S}}%
^{(p)}\text{.}
\end{equation}

\subsection{Energy density and temperature}

The balance equation for total energy density is%
\begin{equation}
\frac{\partial }{\partial t}E+\overrightarrow{\nabla }\cdot \left( 
\overrightarrow{u}E\right) +\overrightarrow{\nabla }\cdot \overrightarrow{q}+%
\overrightarrow{\nabla }\cdot \left( \overrightarrow{u}\cdot 
\overleftrightarrow{P}\right) +\overrightarrow{\nabla }\cdot \left( \frac{1}{%
2}u^{2}\overrightarrow{Q}\right) =\xi +\overrightarrow{u}\cdot \overline{%
\overrightarrow{S}}^{(p)}+\frac{1}{2}u^{2}\overline{S}^{(\rho )}.
\end{equation}%
where the new source term, arising if energy is not conserved by the
collisions, is%
\begin{eqnarray}
\xi \left( \overrightarrow{r},t\right) &=&\frac{1}{2}\sum_{l_{1}l_{2}l_{1}^{%
\prime }l_{2}^{\prime }}\int dx_{1}dx_{2}\;\left( \overrightarrow{q}%
_{12}\cdot \overrightarrow{v}_{12}\right) \delta \left( q_{12}-\sigma
_{l_{1}l_{2}}\right) \Theta \left( -\widehat{q}_{12}\cdot \overrightarrow{v}%
_{12}\right) \\
&&\times \left[ \overline{\delta E_{l_{1}l_{2}}^{l_{1}^{\prime
}l_{2}^{\prime }}}+\frac{1}{2}\delta m_{l_{1}l_{2}}^{l_{1}^{\prime
}l_{2}^{\prime }}\left( \overrightarrow{V}-\overrightarrow{u}\right) ^{2}%
\right] f_{l_{1}l_{2}}\left( x_{1},x_{2};t\right)
K_{l_{1}l_{2}}^{l_{1}^{\prime }l_{2}^{\prime }}\left( x_{12}\right) \delta
\left( \overrightarrow{r}-\overrightarrow{q}_{1}\right) ,  \notag
\end{eqnarray}%
which is recognized as the generalization of the source term studied in the
context of granular fluids. The heat flux is written as a sum of several
contributions%
\begin{equation}
\overrightarrow{q}=\overrightarrow{q}^{K}+\overrightarrow{q}^{V}+%
\overrightarrow{q}^{m}+\overrightarrow{q}^{\delta E}
\end{equation}%
where the kinetic part has the usual form%
\begin{equation*}
\overrightarrow{q}^{K}\left( \overrightarrow{r},t\right) =\sum_{l}\frac{1}{2}%
m_{l}\int d\overrightarrow{v}_{1}\;f_{l}\left( \overrightarrow{r},%
\overrightarrow{v}_{1},t\right) \overrightarrow{V}_{1}V_{1}^{2}
\end{equation*}%
as does the first part of the collisional contribution%
\begin{eqnarray}
\overrightarrow{q}^{V}\left( \overrightarrow{r},t\right) &=&-\frac{1}{2}%
\sum_{l_{1}l_{2}l_{1}^{\prime }l_{2}^{\prime }}\int dx_{1}dx_{2}\;%
\overrightarrow{q}_{12}\left( \overrightarrow{q}_{12}\cdot \overrightarrow{v}%
_{12}\right) \delta \left( q_{12}-\sigma _{l_{1}l_{2}}\right) \Theta \left( -%
\widehat{q}_{12}\cdot \overrightarrow{v}_{12}\right) \\
&&\times f_{l_{1}l_{2}}\left( x_{1},x_{2};t\right)
K_{l_{1}l_{2}}^{l_{1}^{\prime }l_{2}^{\prime }}\left( x_{12}\right) \left( 
\overrightarrow{V}-\overrightarrow{u}\right) \cdot \widetilde{%
\overrightarrow{\gamma }}_{l_{1}l_{2}}^{l_{1}^{\prime }l_{2}^{\prime
}}\int_{0}^{1}dx\;\delta \left( \overrightarrow{r}-x\overrightarrow{q}%
_{1}-\left( 1-x\right) \overrightarrow{q}_{2}\right) ,  \notag
\end{eqnarray}%
which is a measure of energy displacement during the collision (i.e., one
atom experiences a net gain of energy, the other a net loss and this
represents an instantaneous movement of energy from the location of the
second atom to the location of the first). Qualitatively new contributions
arise from the instantaneous transfer of mass,%
\begin{eqnarray}
\overrightarrow{q}^{m}\left( \overrightarrow{r},t\right) &=&-\frac{1}{2}%
\sum_{l_{1}l_{2}l_{1}^{\prime }l_{2}^{\prime }}\frac{m_{l_{2}^{\prime
}}m_{l_{1}}-m_{l_{1}^{\prime }}m_{l_{2}}}{\left( m_{l_{2}^{\prime
}}+m_{l_{1}^{\prime }}\right) \left( m_{l_{2}}+m_{l_{1}}\right) } \\
&&\times \int dx_{1}dx_{2}\;\overrightarrow{q}_{12}\left( \overrightarrow{q}%
_{12}\cdot \overrightarrow{v}_{12}\right) \delta \left( q_{12}-\sigma
_{l_{1}l_{2}}\right) \Theta \left( -\widehat{q}_{12}\cdot \overrightarrow{v}%
_{12}\right)  \notag \\
&&\times f_{l_{1}l_{2}}\left( x_{1},x_{2};t\right)
K_{l_{1}l_{2}}^{l_{1}^{\prime }l_{2}^{\prime }}\left( x_{12}\right) \mu
_{l_{1}l_{2}}v^{2}  \notag \\
&&\times \int_{0}^{1}dx\;\delta \left( \overrightarrow{r}-x\overrightarrow{q}%
_{1}-\left( 1-x\right) \overrightarrow{q}_{2}\right) ,  \notag
\end{eqnarray}%
and from the loss of energy,%
\begin{eqnarray}
\overrightarrow{q}^{\delta E}\left( \overrightarrow{r},t\right) &=&-\frac{1}{%
2}\sum_{l_{1}l_{2}l_{1}^{\prime }l_{2}^{\prime }}\frac{m_{l_{1}^{\prime
}}-m_{l_{2}^{\prime }}}{m_{l_{1}^{\prime }}+m_{l_{2}^{\prime }}} \\
&&\times \int dx_{1}dx_{2}\;\overrightarrow{q}_{12}\left( \overrightarrow{q}%
_{12}\cdot \overrightarrow{v}_{12}\right) \delta \left( q_{12}-\sigma
_{l_{1}l_{2}}\right) \Theta \left( -\widehat{q}_{12}\cdot \overrightarrow{v}%
_{12}\right)  \notag \\
&&\times f_{l_{1}l_{2}}\left( x_{1},x_{2};t\right)
K_{l_{1}l_{2}}^{l_{1}^{\prime }l_{2}^{\prime }}\left( x_{12}\right) 
\overline{\delta E_{l_{1}l_{2}}^{l_{1}^{\prime }l_{2}^{\prime }}}  \notag \\
&&\times \int_{0}^{1}dx\;\delta \left( \overrightarrow{r}-x\overrightarrow{q}%
_{1}-\left( 1-x\right) \overrightarrow{q}_{2}\right) .  \notag
\end{eqnarray}%
Alternatively, noting the relation between the total energy and the kinetic
temperature%
\begin{equation}
E=\frac{D}{2}nk_{B}T+\frac{1}{2}\rho u^{2}
\end{equation}%
the evolution of the kinetic temperature is found to be given by%
\begin{equation}
\left( \frac{\partial }{\partial t}+\overrightarrow{u}\cdot \overrightarrow{%
\nabla }\right) T-\frac{T}{n}\overrightarrow{\nabla }\cdot \sum_{l}%
\overrightarrow{j}_{l}^{K}+\frac{2}{Dnk_{B}}\left[ \overleftrightarrow{P}:%
\overrightarrow{\nabla }\overrightarrow{u}+\overrightarrow{\nabla }\cdot 
\overrightarrow{q}\right] =\frac{2}{Dnk_{B}}\xi .
\end{equation}

\subsection{Enskog Approximation}

The expressions for the balance equations are exact. As a consequence, they
depend on both the exact one-body and two-body distribution functions which
are, in principle, determined by the BBGKY hierarchy. For example the
equation for the one-body distribution is explicitly%
\begin{eqnarray}
\left( \frac{\partial }{\partial t}+\overrightarrow{v}_{1}\cdot \frac{%
\partial }{\partial \overrightarrow{q}_{1}}\right) f_{l_{1}}\left(
x_{1};t\right)  &=&-\sum_{a,b,l_{2}}\int d\overrightarrow{q}_{2}d%
\overrightarrow{v}_{2}  \label{BBGKY1} \\
&&\times \left[ \left| \frac{\partial \left( \widehat{b}%
_{ab}^{l_{1}l_{2}}x_{1},\widehat{b}_{ab}^{l_{1}l_{2}}x_{2}\right) }{\partial
\left( x_{1},x_{2}\right) }\right| ^{-1}\left( \widehat{b}%
_{ab}^{l_{1}l_{2}}\right) ^{-1}K_{ab}^{l_{1}l_{2}}\left( x_{12}\right)
-\delta _{l_{1}a}\delta _{l_{2}b}\right]   \notag \\
&&\times \Theta \left( -\overrightarrow{v}_{12}\cdot \overrightarrow{q}%
_{12}\right) \delta \left( q_{12}-\sigma _{l_{1}l_{2}}\right) 
\overrightarrow{v}_{12}\cdot \widehat{q}_{12}  \notag \\
&&\times f_{l_{1}l_{2}}\left( \overrightarrow{q}_{1},\overrightarrow{v}_{1},%
\overrightarrow{q}_{2},\overrightarrow{v}_{2};t\right)   \notag
\end{eqnarray}%
However, since the latter cannot be solved exactly, except in the special
case of equilibrium, it is necessary to introduce an approximation. The most
common approximation is to assume that the velocities of two colliding atoms
are uncorrelated prior to the collision (they are of course correlated after
the collision since the collision itself generates correlations). That this
approximation is sufficient to decouple the BBGKY hierarchy is seen from the
fact that the right hand side of eq. (\ref{BBGKY1}) since the step function, 
$\Theta \left( -\overrightarrow{v}_{12}\cdot \overrightarrow{q}_{12}\right) $%
, is non-zero only for atoms approaching one another and the delta function
restricts the domain to the instant of contact. Thus, the assumption that
atoms are uncorrelated just prior to a collision, Boltzmann's ''assumption
of molecular chaos'', is precisely the statement that%
\begin{eqnarray}
&&\Theta \left( -\overrightarrow{v}_{12}\cdot \overrightarrow{q}_{12}\right)
\delta \left( q_{12}-\sigma _{l_{1}l_{2}}\right) f_{l_{1}l_{2}}\left( 
\overrightarrow{q}_{1},\overrightarrow{v}_{1},\overrightarrow{q}_{2},%
\overrightarrow{v}_{2}\right)   \label{mc} \\
&\simeq &\Theta \left( -\overrightarrow{v}_{12}\cdot \overrightarrow{q}%
_{12}\right) \delta \left( q_{12}-\sigma _{l_{1}l_{2}}\right) g\left( 
\overrightarrow{q}_{1},\overrightarrow{q}_{2};t\right) f_{l_{1}}\left(
x_{1};t\right) f_{l_{2}}\left( x_{2};t\right)   \notag
\end{eqnarray}%
which, when substituted into Eq.(\ref{BBGKY1}) gives the Enskog
approximation to the one-body distribution function. The factor of $g\left( 
\overrightarrow{q}_{1},\overrightarrow{q}_{2};t\right) $, the spatial pair
distribution function, allows for spatial correlations which always exist.
In the Revised Enskog Theory, it is approximated by the equilibrium
functional of the density evaluated for the local density field of the fluid%
\cite{RET}. The same approximation can be above to give the corresponding
Enskog-approximation to the balance equations. A final consequence follows
from the second equation of the BBGKY hierarchy which has the form%
\begin{eqnarray}
&&\left( \frac{\partial }{\partial t}+\overrightarrow{v}_{1}\cdot \frac{%
\partial }{\partial \overrightarrow{q}_{1}}+\overrightarrow{v}_{2}\cdot 
\frac{\partial }{\partial \overrightarrow{q}_{2}}+\overline{T}%
_{-}(12)\right) f_{l_{1}l_{2}}\left( x_{1},x_{2};t\right)   \label{BBGKY2} \\
&=&-n\int d3\;\left( \overline{T}_{-}(13)+\overline{T}_{-}(23)\right)
f_{l_{1}l_{2}l_{3}}\left( x_{1},x_{2},x_{3};t\right) .  \notag
\end{eqnarray}%
Since atoms cannot interpenetrate the two-body distribution must have the
form $f_{l_{1}l_{2}}\left( x_{1},x_{2};t\right) =\Theta \left( q_{12}-\sigma
_{l_{1}l_{2}}\right) y_{l_{1}l_{2}}\left( x_{1},x_{2};t\right) $ for some
function $y_{l_{1}l_{2}}\left( x_{1},x_{2};t\right) $ which is continuous at 
$q_{12}=\sigma _{l_{1}l_{2}}$. Then, one expects that the singular terms in
Eq.(\ref{BBGKY2}), arising from the gradient acting on the step function and
from the definition of $\overline{T}_{-}(12)$, must cancel, gives the
constraint%
\begin{equation}
\overrightarrow{v}_{12}\cdot \widehat{q}_{12}\delta \left( q_{12}-\sigma
_{l_{1}l_{2}}\right) f_{l_{1}l_{2}}\left( x_{1},x_{2};t\right) =-\overline{T}%
_{-}(12)f_{l_{1}l_{2}}\left( x_{1},x_{2};t\right) 
\end{equation}%
and some rearrangement, together with the approximation of Eq.(\ref{BBGKY2})
gives%
\begin{eqnarray}
\delta \left( q_{12}-\sigma _{l_{1}l_{2}}\right) f_{l_{1}l_{2}}\left(
x_{1},x_{2};t\right)  &\simeq &\delta \left( q_{12}-\sigma
_{l_{1}l_{2}}\right) g\left( \overrightarrow{q}_{1},\overrightarrow{q}%
_{2};t\right) f_{l_{1}}\left( x_{1};t\right) f_{l_{2}}\left( x_{2};t\right) 
\\
&&-\left( \overrightarrow{v}_{12}\cdot \widehat{q}_{12}\right) ^{-1}%
\overline{T}_{-}(12)g\left( \overrightarrow{q}_{1},\overrightarrow{q}%
_{2};t\right) f_{l_{1}}\left( x_{1};t\right) f_{l_{2}}\left( x_{2};t\right) 
\notag
\end{eqnarray}%
which expresses the two-body distribution function at contact in terms of a
completely uncorrelated piece, the first term on the right, and a correction
that takes into account velocity correlations generated by the collision,
the second term on the right. This can be used to evaluate two-body
correlations at the Enskog level of approximation\cite{Lutsko96, Lutsko2001,
LutskoHCS}.

\section{Chapman-Enskog Solution}

In the previous Section, the exact balance equations were developed and the
Enskog approximation introduced. Next, this framework is used to derive the
explicit equations governing the evolution of the hydrodynamic fields by
means of the Chapman-Enskog approximation. As noted in the Introduction,
previous studies of the kinetic theory for reacting systems have often made
the assumption that the chemical reactions are \emph{slow} relative to the
hydrodynamic time-scales. The primary goal here is draw out, and make more
precise, the meaning of this condition by outlining the Chapman-Enskog
procedure under different assumptions about the speed of the chemical
reactions.

Before beginning, note that the phrase ''hydrodynamic fields'' usually
refers to those local fields which are conserved in the long-wavelength
limit (which is to say that their sum over the entire system is conserved).
For a non-reacting fluid of hard-spheres, this means the local partial
number densities, and the momentum and energy densities. For a reacting
fluid, the partial number densities are not conserved and for models of
endo-/exothermic reactions, even the energy is not be conserved. Following
the practice developed in the study of granular fluids (which are
non-reactive but do not conserve energy) it seems natural to expand the
definition of ''hydrodynamic'' fields to include those fields which would be
conserved in the limit of vanishing reaction probabilities. A partial
justification for this is that all of these fields are necessary to develop
a meaningful description of the non-reacting fluid, so one expects that they
must also be included in any description of the reacting fluid (i.e., a
minimal-coupling argument based on continuity of the description with
respect to the control parameters).

The Chapman-Enskog procedure attempts to construct a so-called normal
solution of the Enskog equation which is to say, a solution which is a local
functional of the (exact) hydrodynamic fields and for which all of the space
and time dependence occurs implicitly through those fields\cite{McLennan}.
This implies that the space and time derivatives of the distribution
function can be written in terms of the corresponding derivatives of the
fields and the functional derivative of the distribution with respect to the
fields. In other words, one has%
\begin{equation}
f_{a}\left( \overrightarrow{q}_{1},\overrightarrow{v}_{1},t\right) =f_{a}%
\left[ \overrightarrow{v}_{1}|x_{i}\left( \overrightarrow{q}_{1},t\right)
,n\left( \overrightarrow{q}_{1},t\right) ,\overrightarrow{u}\left( 
\overrightarrow{q}_{1},t\right) ,T\left( \overrightarrow{q}_{1},t\right) %
\right]  \label{norm1}
\end{equation}%
so all of the dependence on $\overrightarrow{q}_{1}$ and $t$ occur through
the hydrodynamic fields so that the time derivative can be expressed as%
\begin{equation}
\frac{\partial }{\partial t}f_{a}\left( \overrightarrow{q}_{1},%
\overrightarrow{v}_{1},t\right) =\sum_{i}\frac{\partial x_{i}}{\partial t}%
\frac{\partial f_{a}}{\partial x_{i}}+\frac{\partial n}{\partial t}\frac{%
\partial f_{a}}{\partial n}+\frac{\partial \overrightarrow{u}}{\partial t}%
\cdot \frac{\partial f_{a}}{\partial \overrightarrow{u}}+\frac{\partial T}{%
\partial t}\frac{\partial f_{a}}{\partial T}.  \label{norm2}
\end{equation}%
Then, the kinetic equation determines the functional dependence of the
distribution on the fields and their derivatives, while the fields are in
turn fixed self-consistently by the balance equations (in the Enskog
approximation).

A further approximation which is made in practical calculations is to assume
that spatial gradients are small so that the equations can be solved
perturbatively via a gradient expansion. To order the terms, one introduces
a uniformity parameter $\epsilon $ and replace $\overrightarrow{\nabla }$
with $\epsilon \overrightarrow{\nabla }$ and order terms in $\epsilon $.
Since the space and time derivatives are related by the balance equations,
one also introduces an expansion of the time derivative $\frac{\partial }{%
\partial t}\equiv \partial _{t}=\partial _{t}^{(0)}+\epsilon \partial
_{t}^{(1)}+...$ as well as of the distribution itself 
\begin{equation}
f_{a}\left( \overrightarrow{q}_{1},\overrightarrow{v}_{1},t\right) =f_{a}^{0}%
\left[ \overrightarrow{v}_{1}|x_{i},n,\overrightarrow{u},T\right] +\epsilon
f_{a}^{1}\left[ \overrightarrow{v}_{1}|x_{i},n,\overrightarrow{u},T\right]
+...
\end{equation}%
where the notation indicates that the distribution is a functional of the
hydrodynamic fields. These expansions are substituted into both the Enskog
equation and the balance equations and an order-by-order solution is sought.
Writing the Enskog equation as%
\begin{equation}
\left( \frac{\partial }{\partial t}+\overrightarrow{v}_{1}\cdot \frac{%
\partial }{\partial \overrightarrow{q}_{1}}\right) f_{a}\left(
x_{1};t\right) =\sum_{bcd}J_{ab,cd}\left[ f_{c},f_{d}\right]
\end{equation}%
so that%
\begin{eqnarray}
J_{ab,cd}\left[ f_{c},f_{d}\right] &=&\int d\overrightarrow{q}_{2}d%
\overrightarrow{v}_{2}\left[ \left| \frac{\partial \left( \widehat{b}%
_{ab}^{cd}x_{1},\widehat{b}_{ab}^{cd}x_{2}\right) }{\partial \left(
x_{1},x_{2}\right) }\right| ^{-1}\left( \widehat{b}_{ab}^{cd}\right)
^{-1}K_{ab}^{cd}\left( x_{12}\right) -\delta _{ac}\delta _{bd}\right] \\
&&\times \Theta \left( -\overrightarrow{v}_{12}\cdot \overrightarrow{q}%
_{12}\right) \delta \left( q_{12}-\sigma _{cd}\right) \overrightarrow{v}%
_{12}\cdot \widehat{q}_{12}f_{c}\left( \overrightarrow{q}_{1},%
\overrightarrow{v}_{1};t\right) f_{d}\left( \overrightarrow{q}_{2},%
\overrightarrow{v}_{2};t\right)  \notag
\end{eqnarray}%
it is also necessary to expand the non-locality of the collision operator
which comes from the term $\delta \left( q_{12}-\sigma _{cd}\right) =\delta
\left( q_{12}\right) +\epsilon \sigma _{cd}\delta ^{\prime }\left(
q_{12}\right) +...$where the derivative of the delta-function, which will
give rise to spatial gradients of the distribution, are scaled with an
appropriate factor of $\epsilon $. In order to control the speed of the
chemistry relative to the hydrodynamics, the non-diagonal part of the
reaction probabilities is separated out as%
\begin{equation}
K_{ab}^{cd}\rightarrow \delta _{ac}\delta _{bd}+\epsilon ^{\alpha }\left(
K_{ab}^{cd}-\delta _{ac}\delta _{bd}\right)  \label{Kelastic}
\end{equation}%
giving%
\begin{equation}
J_{ab,cd}\left[ f_{c},f_{d}\right] =\delta _{ac}\delta
_{bd}J_{a,b}^{(invariat)}\left[ f_{a},f_{b}\right] +\epsilon ^{\alpha
}J_{ab,cd}^{(reactive)}\left[ f_{c},f_{d}\right]
\end{equation}%
where the non-reactive, or invariant, part is the usual collision operator
for non-reactive\ (but possible energy non-conserving) multi-component fluids%
\begin{eqnarray}
J_{a,b}^{(invariant)}\left[ f_{a},f_{b}\right] &=&\int d\overrightarrow{q}%
_{2}d\overrightarrow{v}_{2}\left[ \left| \frac{\partial \left( \widehat{b}%
_{ab}^{ab}x_{1},\widehat{b}_{ab}^{ab}x_{2}\right) }{\partial \left(
x_{1},x_{2}\right) }\right| ^{-1}\left( \widehat{b}_{ab}^{ab}\right) ^{-1}-1%
\right] \\
&&\times \Theta \left( -\overrightarrow{v}_{12}\cdot \overrightarrow{q}%
_{12}\right) \delta \left( q_{12}-\sigma _{ab}\right) \overrightarrow{v}%
_{12}\cdot \widehat{q}_{12}f_{a}\left( \overrightarrow{q}_{1},%
\overrightarrow{v}_{1};t\right) f_{b}\left( \overrightarrow{q}_{2},%
\overrightarrow{v}_{2};t\right)  \notag
\end{eqnarray}%
and, as indicated, the reactive part of the collision operator will be
arbitrarily treated as being of order $\alpha $ in the gradient expansion.
Thus, the full expansion of the collision operator will take the form%
\begin{eqnarray}
J_{ab,cd}\left[ f_{c},f_{d}\right] &=&\delta _{ac}\delta _{bd}\left(
J_{a,b}^{(invariant)0}\left[ f_{a},f_{b}\right] +\epsilon
J_{a,b}^{(invariant)1}\left[ f_{a},f_{b}\right] +...\right) \\
&&+\epsilon ^{\alpha }\left( J_{ab,cd}^{(reactive)0}\left[ f_{c},f_{d}\right]
+\epsilon J_{ab,cd}^{(reactive)1}\left[ f_{c},f_{d}\right] +...\right) . 
\notag
\end{eqnarray}

\subsection{Zeroth-order}

The zeroth order equation for the distribution is then%
\begin{equation}
\partial _{t}^{0}f_{a}^{(0)}=\sum_{b}J_{a,b}^{(elastic)0}\left[
f_{a}^{0},f_{b}^{0}\right] +\delta _{\alpha
0}\sum_{bcd}J_{ab,cd}^{(reactive)0}\left[ f_{c}^{0},f_{d}^{0}\right]
\label{zeroth}
\end{equation}%
which must be supplemented by the corresponding equations for the fields
expanded to zeroth order 
\begin{eqnarray}
\partial _{t}^{0}x_{l} &=&\delta _{\alpha 0}n^{-1}S_{l}^{(n)(reactive)0}
\label{balance0} \\
\partial _{t}^{0}n &=&0  \notag \\
\partial _{t}^{0}\overrightarrow{u} &=&0  \notag \\
\partial _{t}^{0}T &=&\xi ^{(invariant)0}+\delta _{\alpha 0}\xi
^{(reactive)0}.  \notag
\end{eqnarray}%
These balance equations, together with the assumption of normality, eq. (\ref%
{norm2}), serve to define the meaning of the term $\partial
_{t}^{0}f_{a}^{(0)}$ in eq.(\ref{zeroth}). Note that the fluxes do not
enter, being of first order in the gradients, and that the sources are
separated into a non-reactive and reactive part using eq.(\ref{Kelastic}).
For the concentration, mass and velocity fields, there are in general no
non-reactive contributions to the sources whereas for the temperature, there
is the possibility of such a contribution in which case, one recovers the
inelastic hard-sphere system used to model granular fluids. Furthermore, use
has been made of the fact that $f_{a}^{(0)}$ must be a function of $%
\overrightarrow{v}_{1}-\overrightarrow{u}$ which implies that $\overline{%
\overrightarrow{S}}^{(p)(reactive)0}=0$ (since there are no other
zeroth-order vectors) so that no source can appear in the velocity equation,
at this order. These zeroth-order equations illustrate a complication that
occurs for fast reacting systems (e.g. $\alpha =0$) compared to non-reacting
multi-component systems: namely that the sources in the balance equations at
order $n$ require knowledge of the $n$-th order distribution. For
non-reacting elastic systems, the $n$-th order balance equations generally
require only the $n-1$ order distribution so that there is no coupling
between the two. Non-reacting inelastic systems, i.e. granular fluids, share
this complication as can be seen from the appearance of the source term $\xi
^{(non)0}$ in Eqs.(\ref{balance0}).

For $\alpha >0$, only the temperature can have a zeroth-order time
dependence and so can contribute to the left side of eq.(\ref{zeroth}). If
this temperature source is zero then the left hand side of eq.(\ref{zeroth})
is zero and the $f_{a}^{(0)}$ will simply be proportional to a Maxwellian.
The solution of eqs.(\ref{zeroth}) and (\ref{balance0}) for the case $\alpha
>0$ and the non-reactive source in the temperature equation being non-zero
corresponds to the so-called homogeneous cooling state in granular fluids
and has been discussed in detail in the literature for single-component\cite%
{ErnstHCS} and multi-component systems\cite{HCS_Mix}.

\subsection{First-order}

At first order, one has%
\begin{eqnarray}
&&\partial _{t}^{0}f_{a}^{(1)}+\left( \partial _{t}^{1}+\overrightarrow{v}%
_{1}\cdot \overrightarrow{\nabla }\right) f_{a}^{0} \\
&=&\sum_{b}\left( J_{a,b}^{(invariant)0}\left[ f_{a}^{0},f_{b}^{1}\right]
+J_{a,b}^{(invariant)0}\left[ f_{a}^{1},f_{b}^{0}\right]
+J_{a,b}^{(invariant)1}\left[ f_{a}^{0},f_{b}^{0}\right] \right)   \notag \\
&&+\delta _{\alpha 0}\sum_{bcd}\left( J_{ab,cd}^{(reactive)0}\left[
f_{c}^{0},f_{d}^{1}\right] +J_{ab,cd}^{(reactive)0}\left[ f_{c}^{1},f_{d}^{0}%
\right] +J_{ab,cd}^{(reactive)1}\left[ f_{c}^{0},f_{d}^{0}\right] \right)  
\notag \\
&&+\delta _{\alpha 1}\sum_{bcd}J_{ab,cd}^{(reactive)0}\left[
f_{c}^{0},f_{d}^{0}\right]   \notag
\end{eqnarray}%
and for the fields%
\begin{eqnarray}
\partial _{t}^{1}x_{l}+\overrightarrow{u}\cdot \overrightarrow{\nabla }x_{l}
&=&\delta _{\alpha 0}n^{-1}S_{l}^{(n)(reactive)1}  \label{balance1} \\
&&+\delta _{\alpha 1}n^{-1}S_{l}^{(n)(reactive)0}  \notag \\
\partial _{t}^{1}n+\overrightarrow{\nabla }\cdot \left( \overrightarrow{u}%
n\right)  &=&0  \notag \\
\partial _{t}^{1}\overrightarrow{u}+\overrightarrow{u}\cdot \overrightarrow{%
\nabla }\overrightarrow{u}+\rho ^{-1}\overrightarrow{\nabla }p^{(0)}
&=&\delta _{\alpha 0}\rho ^{-1}\overline{\overrightarrow{S}}^{(p)(0)}  \notag
\\
\partial _{t}^{1}T+\overrightarrow{u}\cdot \overrightarrow{\nabla }T+\frac{2%
}{Dnk_{B}}\left[ p^{(0)}\overrightarrow{\nabla }\cdot \overrightarrow{u}+%
\overrightarrow{\nabla }\cdot \left( \overrightarrow{u}w^{(0)}\right) \right]
&=&\xi ^{(invariant)1}  \notag \\
&&+\delta _{\alpha 0}\xi ^{(reactive)1}+\delta _{\alpha 1}\xi ^{(reactive)0}
\notag
\end{eqnarray}%
where we have used the fact that at zeroth order there are no
velocity-independent vectors and only the unit tensor available so that we
must have $\overleftrightarrow{P}^{(0)}=p^{(0)}\overleftrightarrow{1}$, $%
\overleftrightarrow{W}^{(0)}=w^{(0)}\overleftrightarrow{1}$ and all vector
fluxes must vanish.

In general, the first order distribution must take the form%
\begin{equation}
f_{l}^{(1)}\left( \overrightarrow{r},\overrightarrow{V};t\right) =nx_{l}\phi
_{l}\left( \overrightarrow{V}\right) \left[ 
\begin{array}{c}
h_{l}(\overrightarrow{V})+\mathcal{A}_{l}(\overrightarrow{V})\overrightarrow{%
V}\cdot \overrightarrow{\nabla }n+\mathcal{B}_{l}\overrightarrow{V}\cdot 
\overrightarrow{\nabla }T \\ 
+\overleftrightarrow{\mathcal{C}}_{l}:\left( \overrightarrow{\nabla }%
\overrightarrow{V}-\frac{1}{D}\overleftrightarrow{1}\overrightarrow{\nabla }%
\cdot \overrightarrow{V}\right) + \\ 
\mathcal{D}_{l}\overrightarrow{\nabla }\cdot \overrightarrow{V}+\sum_{k}%
\mathcal{E}_{lk}\overrightarrow{V}\cdot \overrightarrow{\nabla }x_{k}%
\end{array}%
\right]  \label{dist1}
\end{equation}%
where I have written the zeroth order distribution in the form $%
f_{l}^{(0)}=nx_{l}\phi _{l}\left( V\right) $. Here the coefficients $%
\mathcal{A}_{l},\mathcal{B}_{l},...$are scalar functions of the velocity
(and in general depend also on space and time through a dependence on the
local hydrodynamic variables as does the zeroth-order distribution, although
for the sake of conciseness this dependence has been suppressed). The
function $h(V)$ represents the first-order correction to $\phi \left(
V\right) $ due to the energy-dependent chemical reactions: for example, if
the only allowed interaction were $A+A\rightarrow A+B$ and this only took
place if the CM kinetic energy were greater than some threshold, $E_{AB}$,
then starting with a system of all $A$-type atoms, one would expect to build
up a preponderance of fast $B$ atoms and a corresponding deficit of fast $A$
atoms. It vanishes in the case that $\alpha >2$ and energy is conserved by
the non-reactive dynamics. The consequences of different orderings of the
reaction terms will be considered separately.

\subsubsection{Ultra-slow reactions:\ $\protect\alpha >2$}

In this case, there are no reactive terms in the first order equations. The
solution is therefore the same as for the equilibrium (or HCS)\
multi-component system. The second order balance equations will also have no
reactive terms. Summing up to second order, the Navier-Stokes order balance
equations are then%
\begin{gather}
\frac{\partial }{\partial t}x_{l}+\overrightarrow{u}\cdot \overrightarrow{%
\nabla }x_{l}+n^{-1}\left[ \overrightarrow{\nabla }\cdot \overrightarrow{j}%
_{l}-x_{l}\overrightarrow{\nabla }\cdot \sum_{l}\overrightarrow{j}_{l}^{K}%
\right] =0  \label{NS} \\
\frac{\partial }{\partial t}n+\overrightarrow{\nabla }\cdot \overrightarrow{u%
}n+\overrightarrow{\nabla }\cdot \sum_{l}\overrightarrow{j}_{l}^{K}=0  \notag
\\
\frac{\partial }{\partial t}\overrightarrow{u}+\overrightarrow{u}\cdot 
\overrightarrow{\nabla }\overrightarrow{u}+\rho ^{-1}\left( \overrightarrow{%
\nabla }\cdot \overleftrightarrow{P}-\overrightarrow{u}\overrightarrow{%
\nabla }\cdot \overrightarrow{Q}\right) =0  \notag \\
\frac{\partial }{\partial t}T+\overrightarrow{u}\cdot \overrightarrow{\nabla 
}T-\frac{T}{n}\overrightarrow{\nabla }\cdot \sum_{l}\overrightarrow{j}%
_{l}^{K}+\frac{2}{Dnk_{B}}\left[ \overleftrightarrow{P}:\overrightarrow{%
\nabla }\overrightarrow{u}+\overrightarrow{\nabla }\cdot \overrightarrow{q}+%
\overrightarrow{\nabla }\cdot \left( \overrightarrow{u}\cdot 
\overleftrightarrow{W}\right) +\frac{1}{2}\overrightarrow{Q}\cdot 
\overrightarrow{\nabla }u^{2}\right] =\xi ^{(invariant)}
\end{gather}

where the fluxes are the sum of zeroth- and first-order contributions,$%
\overleftrightarrow{P}=\overleftrightarrow{P}^{(0)}+\overleftrightarrow{P}%
^{(1)}$, and the source $\xi $ consists of (non-reactive) contributions
summed through second order. This means that for the granular case, $\xi
^{(invariant)}\neq 0$, the Navier-Stokes order balance equations require
knowledge of the second-order (or Burnett order) distribution function.
There is, at this order, no coupling between the hydrodynamic equations and
the reaction equations. Inserted into Eq.(\ref{NS}), the result would be the
Navier-Stokes equations for elastic hard-spheres, or their generalization
for inelastic hard spheres. If this expansion is continued, the $\alpha -th$
order balance equation for the concentrations would be%
\begin{equation}
\partial ^{\left( \alpha \right) }x_{l}+n^{-1}\left[ \overrightarrow{\nabla }%
\cdot \overrightarrow{j}_{l}^{(\alpha -1)}-x_{l}\overrightarrow{\nabla }%
\cdot \sum_{l}\overrightarrow{j}_{l}^{K(\alpha -1)}\right]
=n^{-1}S_{l}^{(n)(reactive)0}.
\end{equation}%
Clearly, the reaction equation remains unknown in this case since one would
need to consistently include the higher order hydrodynamic contributions
that would come from the number current which, in turn, would bring in
couplings to higher-order gradients of the hydrodynamic fields. Without
knowledge of these higher order terms (and they are not known for even the
one-component fluid) the reaction equation can only be consistently studied
in the \emph{absence of hydrodynamic gradients} when the reactive terms are
treated as of order $\alpha >2$.

\subsubsection{Slow reactions:$\;\protect\alpha =2$}

In this case, the first order solution is again the same as in the
non-reacting case. However, the sources will have second-order contributions
so that the Navier-Stokes equations take the form%
\begin{gather}
\frac{\partial }{\partial t}x_{l}+\overrightarrow{u}\cdot \overrightarrow{%
\nabla }x_{l}+n^{-1}\left[ \overrightarrow{\nabla }\cdot \overrightarrow{j}%
_{l}-x_{l}\overrightarrow{\nabla }\cdot \sum_{l}\overrightarrow{j}_{l}^{K}%
\right] =n^{-1}S_{l}^{(n)(reactive)0}  \label{phenom} \\
\frac{\partial }{\partial t}n+\overrightarrow{\nabla }\cdot \overrightarrow{u%
}n+\overrightarrow{\nabla }\cdot \sum_{l}\overrightarrow{j}_{l}^{K}=0  \notag
\\
\frac{\partial }{\partial t}\overrightarrow{u}+\overrightarrow{u}\cdot 
\overrightarrow{\nabla }\overrightarrow{u}+\rho ^{-1}\left( \overrightarrow{%
\nabla }\cdot \overleftrightarrow{P}-\overrightarrow{u}\overrightarrow{%
\nabla }\cdot \overrightarrow{Q}\right) =\rho ^{-1}\overline{\overrightarrow{%
S}}^{(p)(reactive)0}  \notag \\
\frac{\partial }{\partial t}T+\overrightarrow{u}\cdot \overrightarrow{\nabla 
}T-\frac{T}{n}\overrightarrow{\nabla }\cdot \sum_{l}\overrightarrow{j}%
_{l}^{K}+\frac{2}{Dnk_{B}}\left[ \overleftrightarrow{P}:\overrightarrow{%
\nabla }\overrightarrow{u}+\overrightarrow{\nabla }\cdot \overrightarrow{q}+%
\overrightarrow{\nabla }\cdot \left( \overrightarrow{u}\cdot 
\overleftrightarrow{W}\right) +\frac{1}{2}\overrightarrow{Q}\cdot 
\overrightarrow{\nabla }u^{2}\right]   \notag \\
=\xi ^{(invariant)}+\xi ^{(reactive)0}.
\end{gather}%
For the simplest case that the reactions conserve energy and momentum, the
reactions are governed by exactly the convective-reaction-diffusion equation
that one might expect. The reaction rates are calculated using the local
equilibrium distribution as in elementary treatments\cite{McQuarry}. Except
for the usual modification of the transport properties arising from the use
of the Enskog equation, as opposed to the Boltzmann equation, there are no
new dense-fluid effects.

\subsubsection{Moderate reactions: $\protect\alpha =1$}

For moderately fast reactions, the situation becomes more interesting.
Considering here only the case that mass and energy are conserved by all
collisions, the first order balance equations - the generalization of the
Euler equations - are found to be 
\begin{eqnarray}
\left( \frac{\partial }{\partial t}+\overrightarrow{u}\cdot \overrightarrow{%
\nabla }\right) x_{l} &=&n^{-1}S_{l}^{(n)(reactive)0} \\
\frac{\partial }{\partial t}n+\overrightarrow{\nabla }\cdot \overrightarrow{u%
}n &=&0  \notag \\
\left( \frac{\partial }{\partial t}+\overrightarrow{u}\cdot \overrightarrow{%
\nabla }\right) \overrightarrow{u}+\rho ^{-1}\overrightarrow{\nabla }p^{(0)}
&=&0  \notag \\
\left( \frac{\partial }{\partial t}+\overrightarrow{u}\cdot \overrightarrow{%
\nabla }\right) T+\frac{2}{Dnk_{B}}p\overrightarrow{\nabla }\cdot 
\overrightarrow{u} &=&0.  \notag
\end{eqnarray}%
so that the reactions, with reaction rates calculated from the
local-equilibrium distribution function, enter into the Euler equations. The
Navier-Stokes equations will involve the reaction rates calculated up to
first order in the distribution. In general, the only nonzero coupling in
the reaction source will take the form $S_{l}^{(n)(reactive)1}=S_{l}^{(n)}%
\overrightarrow{\nabla }\cdot \overrightarrow{u}$ where $S_{l}^{(n)}$ is a
scalar function of the concentrations, density and temperature. The
Navier-Stokes equations will therefore take the form 
\begin{align}
\left( \frac{\partial }{\partial t}+\overrightarrow{u}\cdot \overrightarrow{%
\nabla }\right) x_{l}+n^{-1}\left( \overrightarrow{\nabla }\cdot 
\overrightarrow{j}_{l}-x_{l}\overrightarrow{\nabla }\cdot \sum_{l}%
\overrightarrow{j}_{l}^{K}\right) & =n^{-1}S_{l}^{(n)(reactive)0} \label{reaction1}\\
& +n^{-1}S_{l}^{(n)(reactive)1}+S_{l}^{(n)}\overrightarrow{\nabla }\cdot 
\overrightarrow{u}  \notag \\
\frac{\partial }{\partial t}n+\overrightarrow{\nabla }\cdot \overrightarrow{u%
}n+\overrightarrow{\nabla }\cdot \sum_{l}\overrightarrow{j}_{l}^{K}& =0 
\notag \\
\left( \frac{\partial }{\partial t}+\overrightarrow{u}\cdot \overrightarrow{%
\nabla }\right) \overrightarrow{u}+\rho ^{-1}\overrightarrow{\nabla }\cdot 
\overleftrightarrow{P}& =0  \notag \\
\left( \frac{\partial }{\partial t}+\overrightarrow{u}\cdot \overrightarrow{%
\nabla }\right) T-\frac{T}{n}\overrightarrow{\nabla }\cdot \sum_{l}%
\overrightarrow{j}_{l}^{K}+\frac{2}{Dnk_{B}}\left( \overleftrightarrow{P}:%
\overrightarrow{\nabla }\overrightarrow{u}+\overrightarrow{\nabla }\cdot 
\overrightarrow{q}\right) & =0.  \notag
\end{align}%
The source term for the reactions has three contributions:\ the zeroth order
reaction rate (calculated using the local equilibrium distribution
function), the first order correction (due to deviations of the distribution
from local equilibrium) and a new, dense-fluid effect which couples the
reactions to the divergence of the velocity field with some field-dependent
coefficient,$S_{l}^{(n)}$ . This coupling is a dense-fluid effect which does
not exist in the Boltzmann approximation and not surprisingly, its origin is
closely related to that of the bulk-viscosity which is also zero in the
Boltzmann theory, but not the Enskog theory. (The calculation of these terms
will be discussed in detail in a future publication but the fact that these
are the only possible couplings is due to the fact that no other
galilean-invariant scalars, linear in the gradients of the fields, can be
constructed).

\subsubsection{Fast reactions:\ $\protect\alpha =0$}

In the case of fast reactions, no a priori assumption is made about the
speed of the reactions compared to the hydrodynamic time scales. The balance
equations to first order, i.e. the Euler equations, are 
\begin{eqnarray}
\left( \frac{\partial }{\partial t}+\overrightarrow{u}\cdot \overrightarrow{%
\nabla }\right) x_{l} &=&n^{-1}S_{l}^{(n)(reactive)0}+S_{l}^{(n)}%
\overrightarrow{\nabla }\cdot \overrightarrow{u} \\
\left( \frac{\partial }{\partial t}+\overrightarrow{u}\cdot \overrightarrow{%
\nabla }\right) n+n\overrightarrow{\nabla }\cdot \overrightarrow{u} &=&0 
\notag \\
\left( \frac{\partial }{\partial t}+\overrightarrow{u}\cdot \overrightarrow{%
\nabla }\right) \overrightarrow{u}+\rho ^{-1}\overrightarrow{\nabla }p &=&0 
\notag \\
\left( \frac{\partial }{\partial t}+\overrightarrow{u}\cdot \overrightarrow{%
\nabla }\right) T+\frac{2}{Dnk_{B}}\left[ p\overrightarrow{\nabla }\cdot 
\overrightarrow{u}+\overrightarrow{\nabla }\cdot \left( w\overrightarrow{u}%
\right) \right] &=&\xi ^{(invariant)0}+\xi ^{(reactive)0}+\xi
^{(invariant)1}+\xi ^{(reactive)1}  \notag
\end{eqnarray}%
so that even the Euler equations show the dense-fluid correction to the
reaction rates. The second-order, or Navier-Stokes, equations require the
evaluation of the source terms to second order, which in turn requires
knowledge of the distribution function to second order (also called Burnett
order). One then expects that even for mass and energy conserving
interactions, the reaction equation will contain couplings to gradients of
all of the hydrodynamic fields. However, since the complete Burnett-order
Chapman-Enskog solution of the Enskog equation is not even known for the
case of a single-component fluid, there is no practical value in continuing
the analysis for this case.

\section{Conclusion}

In this paper, the kinetic theory of reactive hard-core systems has been
extended to include the possibility of mass transfer and/or loss and energy
gain/loss. When mass is not conserved, the collision rule becomes dependent
on the model used to describe the lost mass. Nevertheless, quite general
expressions for the dynamics of phase functions and the distribution
function of the system can be given and used to derive equally general
expressions for the exact balance laws for mass, momentum, energy and
concentration. For example, when mass is conserved but a fixed fraction of
the rest-frame kinetic energy is lost during collisions, the usual inelastic
hard sphere kinetic theory, used as a model of granular fluids\cite%
{HCSLiouville}, is recovered. The kinetic theory was used, within the Enskog
approximation, to discuss the various phenomenological laws, extensions of
the Navier-Stokes equations, that arise from different orderings of the
reaction terms within the Chapman-Enskog procedure. It was noted that the
intuitive model of the Navier-Stokes equations coupled to a
reaction-diffusion equation through a convective term only arises when the
''speed'' of the chemical reactions is comparable to some hydrodynamic time
scale and even in this case, an additional coupling to the divergence of the
velocity field can occur in dense fluids.

How fast do we expect the chemistry to be relative to the hydrodynamics?\ In
the model considered here, chemical reactions cannot be faster than the
collision time. In fact, a typical reaction rate would be something like $%
p\left( \delta x\right) e^{-E/k_{B}T}\nu _{col}$ where $p$ is the
probability of a reaction occuring if the colliding atoms have energy
greater than the reaction energy barrier, $E$, $\delta x$ is the difference
between the concentration of the species and its equilibrium concentration
and $\nu _{col}$ is the collision frequency. On the other hand, the
Chapman-Enskog procedure is based on a gradient expansion, the small
parameter $\epsilon $ will generally be a measure of the ratio of the
typical microscopic length scale, the mean free path $l_{mfp},$ to a typical
length scale for hydrodynamic gradients $L$. (In Fourier space, where
gradients $\overrightarrow{\nabla }$ correspond to wavevectors $%
\overrightarrow{k}$, this becomes $\epsilon \sim kl_{mfp}$.) So, setting $%
pe^{-E/k_{B}T}\nu _{col}\sim \left( kl_{mfp}\right) ^{\alpha }\nu _{col}\sim
\left( l_{mfp}/L\right) ^{\alpha }\nu _{col}$ gives 
\begin{equation}
\alpha \sim \frac{\ln p\left( \delta x\right) -\frac{E}{k_{B}T}}{\ln \left(
l_{mfp}/L\right) }.
\end{equation}%
For systems in which hydrodynamics is applicable, one has $l_{mfp}/L<<1$ so
that $\alpha $ ranges from a minumum of $\frac{\ln p\left( \delta x\right) }{%
\ln \left( l_{mfp}/L\right) }\geq 0$, for $k_{B}T\gg E$, to very large
values for low temperatures. Far from chemical equilibrium, $\delta x\sim 1$%
, the lower limit could be arbitrarily close to zero depending on the
reaction probability so that ''moderate'' and ''fast'' reactions are
possible at high temperatures. Indeed, if all of these parameters are fixed,
then fast reactions will always occur in the hydodrynamic regime limit $%
l_{mfp}/L\rightarrow 0$. The conclusion is that, unless the concentrations
are close to their equilibrium values, the reaction probabilities are very
small or the temperature is extremely low, the concept of slow chemical
reactions may be of limited applicability and, so, the correct
phenomenological description, from the standpoint of kinetic theory, may be
more complex than the reaction-diffusion-advection model.

In summary,if chemical reactions are slow compared to the rate of
dissipation in the fluid, then hydrodynamics and chemistry are not
meaningfully coupled. If the reaction rate is comparable to the rate of
dissipation in the fluid, i.e. $\lambda k^{2}$ for some transport
coefficient $\lambda $ and wavevector $k$, then the usual
reaction-diffusion-advection equation results. For faster reactions,
additional couplings occur and the chemistry and hydrodynamics become more
interdependent. The detailed solution of the Enskog, and the resulting
phenomenological equations for particular reaction models will be the
subject of a future publication where the importance for sonochemistry of
additional terms, such as those occurring in Eq.(\ref{reaction1}), will be
investigated.

\bigskip

\begin{acknowledgments}
The work presented here has benefited from discussions with Jim Dufty, Ray
Kapral, Jean-Pierre Boon and Pierre Gaspard. Some financial support was
received from the Universit\'{e} Libre de Bruxelles.
\end{acknowledgments}

\bigskip

\appendix

\section{\protect\bigskip The hard-core Liouville operator}

\label{AppT}

The goal in this appendix is to provide motivation for the statement in the
text that the form of the pseudo-Liouville operator is independent of the
collision rule. To start with, restrict attention to a system of 2 atoms.
Let $X\left( \Gamma ;t\right) $ be the characteristic function for
collisions after at time $t$ beginning with the phase $\Gamma $ at time $0$
so that if the two atoms do \emph{not }collide between during the interval $%
[0,t]$ then $X\left( \Gamma ;t\right) =0$ whereas if they \emph{do} collide, 
$X\left( \Gamma ;t\right) =1$. Then the time evolution of the phase function 
$A_{\Gamma }\left( t\right) =A\left( \Gamma (t),t\right) $ is given by%
\begin{equation}
A_{\Gamma }\left( t\right) =\left( 1-X\left( \Gamma ;t\right) \right)
A\left( \Gamma _{0}(t),t\right) +X\left( \Gamma ;t\right) A\left( \Gamma
^{\prime }(t),t\right)
\end{equation}%
where $\Gamma _{0}(t)$ is just the phase of the system propagated a time $t$
into the future in the absence of interactions and is explicitly $\Gamma
_{0}(t)=\left( \overrightarrow{q}_{1}+\overrightarrow{v}_{1}t,%
\overrightarrow{v}_{1},\widehat{l}_{1},\overrightarrow{q}_{2}+%
\overrightarrow{v}_{2}t,\overrightarrow{v}_{2},\widehat{l}_{2}\right) $.
(Note that attention is restricted to the case that velocities are constant
during free-streaming:\ generalization to include one-body forces is
straightforward.) The phase point $\Gamma ^{\prime }(t)$ is the position the
system would reach in phase space if a collision occurred at some time $\tau
\left( \Gamma \right) $ $\in $ $\left[ 0,t\right] $. Explicit expressions
can also be given for its components such as $\overrightarrow{q}_{1}^{\prime
}(t)=\overrightarrow{q}_{1}+\overrightarrow{v}_{1}\tau +\overrightarrow{v}%
_{1}^{\prime }\left( t-\tau \right) $, etc. Direct differentiation then gives%
\begin{eqnarray}
\frac{dA_{\Gamma }\left( t\right) }{dt} &=&\frac{\partial A}{\partial t}%
+\left( 1-X\left( \Gamma ;t\right) \right) \frac{\partial q(t)}{\partial t}%
\cdot \frac{\partial }{\partial q(t)}A\left( \Gamma _{0}(t),t\right) \\
&&+X\left( \Gamma ;t\right) \frac{\partial q(t)}{\partial t}\cdot \frac{%
\partial }{\partial q(t)}A\left( \Gamma ^{\prime }(t),t\right)  \notag \\
&&+\frac{dX\left( \Gamma ;t\right) }{dt}\left( A\left( \Gamma ^{\prime
}(t),t\right) -A\left( \Gamma _{0}(t),t\right) \right) .  \notag
\end{eqnarray}%
Now, from the definitions above%
\begin{eqnarray}
&&\left( 1-X\left( \Gamma ;t\right) \right) \frac{\partial q(t)}{\partial t}%
\cdot \frac{\partial }{\partial q(t)}A\left( \Gamma _{0}(t),t\right)
+X\left( \Gamma ;t\right) \frac{\partial q(t)}{\partial t}\cdot \frac{%
\partial }{\partial q(t)}A\left( \Gamma ^{\prime }(t),t\right) \\
&=&\left( 1-X\left( \Gamma ;t\right) \right) \sum_{i=1,2}\overrightarrow{v}%
_{i}\cdot \frac{\partial }{\partial \overrightarrow{q}_{i}(t)}A\left( \Gamma
_{0}(t),t\right) +X\left( \Gamma ;t\right) \sum_{i=1,2}\overrightarrow{v}%
_{i}^{\prime }\cdot \frac{\partial }{\partial \overrightarrow{q}_{i}(t)}%
A\left( \Gamma ^{\prime }(t),t\right)  \notag \\
&=&\sum_{i=1,2}\overrightarrow{v}_{i}(t)\cdot \frac{\partial }{\partial 
\overrightarrow{q}_{i}(t)}A\left( \Gamma (t),t\right)  \notag
\end{eqnarray}%
giving%
\begin{equation}
\frac{dA_{\Gamma }\left( t\right) }{dt}=\frac{\partial A}{\partial t}%
+\sum_{i=1,2}\overrightarrow{v}_{i}(t)\cdot \frac{\partial }{\partial 
\overrightarrow{q}_{i}(t)}A_{\Gamma }\left( t\right) +\frac{dX\left( \Gamma
;t\right) }{dt}\left( A\left( \Gamma ^{\prime }(t),t\right) -A\left( \Gamma
_{0}(t),t\right) \right)
\end{equation}%
Now, since $X\left( \Gamma ;t\right) $ has the form of a step function (it
is zero if $t<\tau \left( \Gamma \right) $ and one otherwise), we must have%
\begin{equation}
\frac{dX\left( \Gamma ;t\right) }{dt}=\delta \left( t-\tau \left( \Gamma
\right) \right) ,  \label{delta}
\end{equation}%
and this also gives the correct result (zero) if $\tau \left( \Gamma \right) 
$ is imaginary (indicating that no collision ever occurs starting from the
given state). Then, using%
\begin{eqnarray}
\delta \left( t-\tau \left( \Gamma \right) \right) \left( A\left( \Gamma
^{\prime }(t),t\right) -A\left( \Gamma _{0}(t),t\right) \right) &=&\delta
\left( t-\tau \left( \Gamma \right) \right) \left( A\left( \Gamma ^{\prime
}(\tau \left( \Gamma \right) ),\tau \left( \Gamma \right) \right) -A\left(
\Gamma _{0}(\tau \left( \Gamma \right) ),\tau \left( \Gamma \right) \right)
\right) \\
&=&\delta \left( t-\tau \left( \Gamma \right) \right) \left(
\sum_{l_{1}^{\prime }l_{2}^{\prime }}\widehat{M}_{l_{1}l_{2}}^{l_{1}^{\prime
}l_{2}^{\prime }}\widehat{b}_{l_{1}l_{2}}^{l_{1}^{\prime }l_{2}^{\prime
}}-1\right) A_{\Gamma }\left( \tau \left( \Gamma \right) \right)  \notag \\
&=&\delta \left( t-\tau \left( \Gamma \right) \right) \left(
\sum_{l_{1}^{\prime }l_{2}^{\prime }}\widehat{M}_{l_{1}l_{2}}^{l_{1}^{\prime
}l_{2}^{\prime }}\widehat{b}_{l_{1}l_{2}}^{l_{1}^{\prime }l_{2}^{\prime
}}-1\right) A_{\Gamma }\left( t\right)  \notag
\end{eqnarray}%
gives%
\begin{equation}
\frac{dA_{\Gamma }\left( t\right) }{dt}=\left[ \frac{\partial }{\partial t}%
+\sum_{i=1,2}\overrightarrow{v}_{i}(t)\cdot \frac{\partial }{\partial 
\overrightarrow{q}_{i}(t)}+\delta \left( t-\tau \left( \Gamma \right)
\right) \left( \sum_{l_{1}^{\prime }l_{2}^{\prime }}\widehat{M}%
_{l_{1}l_{2}}^{l_{1}^{\prime }l_{2}^{\prime }}\widehat{b}%
_{l_{1}l_{2}}^{l_{1}^{\prime }l_{2}^{\prime }}-1\right) \right] A_{\Gamma
}\left( t\right)
\end{equation}%
I order to express the right hand side entirely in terms of $\Gamma (t)$
rather than the initial condition $\Gamma $, the temporal delta-function is
rewritten using 
\begin{equation}
\delta \left( q_{12}\left( t\right) -\sigma _{l_{1}l_{2}}\right) =\sum_{i}%
\frac{\delta \left( t-\tau _{i}\left( \Gamma \right) \right) }{\left| \frac{%
\partial }{\partial t}q_{12}\left( t\right) \right| _{t=\tau _{i}\left(
\Gamma \right) }}
\end{equation}%
where $\tau _{i}\left( \Gamma \right) $ are the roots of $q_{12}^{2}\left(
\tau _{i}\right) -\sigma _{l_{1}l_{2}}^{2}=0$. They correspond to the time
at which the two atoms are first in contact, i.e. the physical collision
time which is denoted as $\tau \left( \Gamma \right) $, and the time at
which they are last in contact if they are allowed to pass through one
another (which is not physical). One picks out the correct root by noting
that at the physical collision time $\overrightarrow{q}_{12}\left( \tau
\right) \cdot \overrightarrow{v}_{12}(\tau )<0$ while the sign is reversed
at the unphysical collision time so that%
\begin{equation}
\delta \left( q_{12}\left( t\right) -\sigma _{l_{1}l_{2}}\right) \Theta
\left( -\overrightarrow{q}_{12}\left( t\right) \cdot \overrightarrow{v}%
_{12}(t)\right) =\frac{\delta \left( t-\tau \left( \Gamma \right) \right) }{%
\left| \frac{\partial }{\partial t}q_{12}\left( t\right) \right| _{t=\tau
\left( \Gamma \right) }}=\frac{\delta \left( t-\tau \left( \Gamma \right)
\right) }{\left| \widehat{q}_{12}\left( \tau \right) \cdot \overrightarrow{v}%
_{12}(\tau )\right| }=\frac{\delta \left( t-\tau \left( \Gamma \right)
\right) }{\left| \widehat{q}_{12}\left( t\right) \cdot \overrightarrow{v}%
_{12}(t)\right| }
\end{equation}%
or%
\begin{equation}
\delta \left( t-\tau \left( \Gamma \right) \right) =\delta \left(
q_{12}\left( t\right) -\sigma _{l_{1}l_{2}}\right) \Theta \left( -%
\overrightarrow{q}_{12}\left( t\right) \cdot \overrightarrow{v}%
_{12}(t)\right) \left| \widehat{q}_{12}\left( t\right) \cdot \overrightarrow{%
v}_{12}(t)\right|
\end{equation}%
giving finally%
\begin{equation}
\frac{dA_{\Gamma }\left( t\right) }{dt}=\left[ \frac{\partial }{\partial t}+%
\widehat{L}(t)\right] A_{\Gamma }\left( t\right)  \label{L1}
\end{equation}%
with%
\begin{equation}
\widehat{L}(t)=\sum_{1\leq i\leq 2}\overrightarrow{v}_{i}(t)\cdot \frac{%
\partial }{\partial \overrightarrow{q}_{i}(t)}+\sum_{1\leq i<j\leq 2}%
\widehat{T}_{+}(12;t)  \label{L2}
\end{equation}%
where, for arbitrary phase function $B\left( \Gamma ,t\right) $, 
\begin{eqnarray}
\widehat{T}_{+}(12;t)B\left( \Gamma (t),t\right) &=&\delta \left(
q_{12}(t)-\sigma _{l_{1}l_{2}}\right) \Theta \left( -\overrightarrow{q}%
_{12}(t)\cdot \overrightarrow{v}_{12}(t)\right) \left| \widehat{q}%
_{12}(t)\cdot \overrightarrow{v}_{12}(t)\right| \\
&&\times \left( \sum_{l_{1}^{\prime }l_{2}^{\prime }}\widehat{M}%
_{l_{1}l_{2}}^{l_{1}^{\prime }l_{2}^{\prime }}\widehat{b}%
_{l_{1}l_{2}}^{l_{1}^{\prime }l_{2}^{\prime }}-1\right) B\left( \Gamma
(t),t\right) .  \notag
\end{eqnarray}%
For more than two atoms, one simply extends the sums in eq.(\ref{L2}) since,
in a finite system, only binary collisions can occur.

Starting with an initial condition $A_{\Gamma }\left( t\right) =A\left(
\Gamma \right) $, iteration of eq.(\ref{L1}) immediately gives%
\begin{equation}
\left. \frac{d^{n}A_{\Gamma }\left( t\right) }{dt^{n}}\right| _{t=0}=%
\widehat{L}^{n}A\left( \Gamma \right)
\end{equation}%
with $\widehat{L}=\widehat{L}(0)$ which implies that%
\begin{equation}
A_{\Gamma }\left( t\right) =\exp \left( \widehat{L}t\right) A\left( \Gamma
\right)
\end{equation}%
and%
\begin{equation}
\frac{d}{dt}A_{\Gamma }\left( t\right) =\widehat{L}\exp \left( \widehat{L}%
t\right) A\left( \Gamma \right) =\widehat{L}A_{\Gamma }\left( t\right)
\end{equation}%
as claimed in the text.

\section{The adjoint Liouville operator}

\label{AppA}

To derive the adjoint operator, begin with its definition%
\begin{equation}
\int d\Gamma \;B\left( \Gamma \right) L_{+}A\left( \Gamma \right) =\int
d\Gamma \;\left( L_{+}^{A}B\left( \Gamma \right) \right) A\left( \Gamma
\right)
\end{equation}%
or, more explicitly,%
\begin{equation}
\sum_{l_{1},l_{2}...}\int dx_{1}dx_{2}...\;\;B\left( \Gamma \right)
L_{+}A\left( \Gamma \right) =\sum_{l_{1},l_{2}}\int dx_{1}dx_{2}\;\;\left(
L_{+}^{A}B\left( \Gamma \right) \right) A\left( \Gamma \right)
\end{equation}%
Now, 
\begin{equation}
L_{+}=L_{+}^{(0)}+\sum_{i<j}T_{+}(ij)
\end{equation}%
and it is obvious that, neglecting surface terms,%
\begin{equation}
\int d\Gamma \;B\left( \Gamma \right) L_{+}^{(0)}A\left( \Gamma \right)
=\int d\Gamma \;\left( -L_{+}^{(0)}B\left( \Gamma \right) \right) A\left(
\Gamma \right)
\end{equation}%
so%
\begin{equation}
L_{+}^{(0)A}=-L_{+}^{(0)}.
\end{equation}

Next, consider one of the collision operators and restrict attention to a
system of two atoms. Then

\begin{eqnarray}
&&\sum_{l_{1},l_{2}}\int dx_{1}dx_{2}\;\;B\left(
x_{1},l_{1};x_{2},l_{2}\right) \left[ T_{+}(12)A\left(
x_{1},l_{1};x_{2},l_{2}\right) \right] \\
&=&-\sum_{l_{1},l_{2},a,b}\int dx_{1}dx_{2}\;B\left(
x_{1},l_{1};x_{2},l_{2}\right) \Theta \left( -\overrightarrow{v}_{12}\cdot 
\overrightarrow{q}_{12}\right) \delta \left( q_{12}-\sigma _{12}\right) 
\overrightarrow{v}_{12}\cdot \overrightarrow{q}_{12}  \notag \\
&&\times \left( K_{l_{1}l_{2}}^{ab}\left( x_{1},l_{1};x_{2},l_{2}\right) A(%
\widehat{b}_{l_{1}l_{2}}^{ab}x_{1},a;\widehat{b}_{l_{1}l_{2}}^{ab}x_{2},b)-%
\delta _{al_{1}}\delta _{bl_{2}}A\left( x_{1},l_{1};x_{2},l_{2}\right)
\right)  \notag
\end{eqnarray}%
Consider the first term. Relabeling the species in the sum gives%
\begin{eqnarray}
&&-\sum_{l_{1},l_{2},a,b}\int dx_{1}dx_{2}\;B\left(
x_{1},l_{1};x_{2},l_{2}\right) \Theta \left( -\overrightarrow{v}_{12}\cdot 
\overrightarrow{q}_{12}\right) \delta \left( q_{12}-\sigma
_{l_{1}l_{2}}\right) \overrightarrow{v}_{12}\cdot \overrightarrow{q}_{12} 
\notag \\
&&\times K_{l_{1}l_{2}}^{ab}\left( x_{1},l_{1};x_{2},l_{2}\right) A(\widehat{%
b}_{l_{1}l_{2}}^{ab}x_{1},a;\widehat{b}_{l_{1}l_{2}}^{ab}x_{2},b) \\
&=&-\sum_{l_{1},l_{2},a,b}\int dx_{1}dx_{2}\;B\left( x_{1},a;x_{2},b\right)
\Theta \left( -\overrightarrow{v}_{12}\cdot \overrightarrow{q}_{12}\right)
\delta \left( q_{12}-\sigma _{ab}\right) \overrightarrow{v}_{12}\cdot 
\overrightarrow{q}_{12}  \notag \\
&&\times K_{ab}^{l_{1}l_{2}}\left( x_{1},a;x_{2},b\right) A(\widehat{b}%
_{ab}^{l_{1}l_{2}}x_{1},l_{1};\widehat{b}_{ab}^{l_{1}l_{2}}x_{2},l_{2}). 
\notag
\end{eqnarray}%
Assuming that the collision operator is invertible, then introducing new
integration variables $y_{i}=\widehat{b}_{ab}^{l_{1}l_{2}}x_{i}$ and the
corresponding Jacobian%
\begin{equation}
J_{ab}^{l_{1}l_{2}}\left( y_{1},y_{2}\right) =\left| \frac{\partial \left(
\left( \widehat{b}_{ab}^{l_{1}l_{2}}\right) ^{-1}y_{1},\left( \widehat{b}%
_{ab}^{l_{1}l_{2}}\right) ^{-1}y_{2}\right) }{\partial \left(
y_{1},y_{2}\right) }\right|
\end{equation}%
gives%
\begin{eqnarray}
&&-\sum_{l_{1},l_{2},a,b}\int dx_{1}dx_{2}\;B\left(
x_{1},l_{1};x_{2},l_{2}\right) \Theta \left( -\overrightarrow{v}_{12}\cdot 
\overrightarrow{q}_{12}\right) \delta \left( q_{12}-\sigma
_{l_{1}l_{2}}\right) \overrightarrow{v}_{12}\cdot \overrightarrow{q}_{12} \\
&&\times K_{l_{1}l_{2}}^{ab}\left( x_{1},l_{1};x_{2},l_{2}\right) A(\widehat{%
b}_{l_{1}l_{2}}^{ab}x_{1},a;\widehat{b}_{l_{1}l_{2}}^{ab}x_{2},b)  \notag \\
&=&-\sum_{l_{1},l_{2},a,b}\int dy_{1}dy_{2}\;J_{ab}^{l_{1}l_{2}}\left(
y_{1},y_{2}\right) B\left( \left( \widehat{b}_{ab}^{l_{1}l_{2}}\right)
^{-1}y_{1},a;\left( \widehat{b}_{ab}^{l_{1}l_{2}}\right) ^{-1}y_{2};b\right)
\Theta \left( -\left( \widehat{b}_{ab}^{l_{1}l_{2}}\right) ^{-1}%
\overrightarrow{v}_{12}\cdot \overrightarrow{q}_{12}\right)  \notag \\
&&\times \delta \left( q_{12}-\sigma _{ab}\right) \left( \left( \widehat{b}%
_{ab}^{l_{1}l_{2}}\right) ^{-1}\overrightarrow{v}_{12}\right) \cdot 
\overrightarrow{q}_{12}K_{ab}^{l_{1}l_{2}}\left( \left( \widehat{b}%
_{ab}^{l_{1}l_{2}}\right) ^{-1}y_{1},a;\left( \widehat{b}_{ab}^{l_{1}l_{2}}%
\right) ^{-1}y_{2},b\right) A(y_{1},l_{1};y_{2},l_{2})  \notag \\
&=&-\sum_{l_{1},l_{2},a,b}\int dy_{1}dy_{2}\;A(y_{1},l_{1};y_{2},l_{2}) 
\notag \\
&&\times \left[ J_{ab}^{l_{1}l_{2}}\left( y_{1},y_{2}\right) \left( \widehat{%
b}_{ab}^{l_{1}l_{2}}\right) ^{-1}\Theta \left( -\overrightarrow{v}_{12}\cdot 
\overrightarrow{q}_{12}\right) \delta \left( q_{12}-\sigma
_{l_{1}l_{2}}\right) \overrightarrow{v}_{12}\cdot \overrightarrow{q}%
_{12}K_{ab}^{l_{1}l_{2}}\left( y_{1},l_{1};y_{2},l_{1}\right) \right]  \notag
\\
&&\times B\left( x_{1},l_{1};x_{2},l_{2}\right)  \notag
\end{eqnarray}%
where the operator $\left( \widehat{b}_{ab}^{l_{1}l_{2}}\right) ^{-1}$ has
the effect of changing the species from $l_{1},l_{2}$ to $a,b$. One can then
write 
\begin{eqnarray}
&&T_{+}^{A}\left( 12\right) B\left( x_{1},l_{1};x_{2},l_{2}\right) \\
&=&-\sum_{ab}\left[ J_{ab}^{l_{1}l_{2}}\left( x_{1},x_{2}\right) \left( 
\widehat{b}_{ab}^{l_{1}l_{2}}\right) ^{-1}K_{ab}^{l_{1}l_{2}}\left(
x_{1},l_{1};x_{2},l_{2}\right) -1\right] \Theta \left( -\overrightarrow{v}%
_{12}\cdot \overrightarrow{q}_{12}\right) \delta \left( q_{12}-\sigma
_{l_{1}l_{2}}\right) \overrightarrow{v}_{12}\cdot \overrightarrow{q}_{12} 
\notag \\
&&\times B\left( x_{1},l_{1};x_{2},l_{2}\right) .  \notag
\end{eqnarray}%
In some cases of interest, the collision dynamics may not be invertible. For
example, suppose that collisions with total rest frame energy less than some
threshold, $E$, are elastic while those with energy greater than this are
inelastic. Then, a pair of atoms with rest frame energy after collision of $%
\frac{1}{2}\mu _{12}v_{12}^{\prime 2}<E$ might have resulted from either (a)
a collision between two atoms with rest-frame energy below the threshold or
(b) a collision between two atoms that had energy above the threshold but
that lost part of this due to the inelastic process. In this case, it is
necessary in the definition of the adjoint operator to include an additional
sum over the various branches of the inverse collision dynamics. Even when
it occurs, such a complication may not be of practical importance since it
is often the case that expressions involving the adjoint operator $T_{+}^{A}$
can be rewritten in terms of the original operator $T_{+}$.

\section{Derivation of the balance equations}

\label{AppB}

In this section, the general form of the local balance equations is derive
and then specialized to the partial density, momentum and energy fields.

\subsection{General form of the balance equations}

Consider any one-body phase function of the form%
\begin{equation}
\widehat{\Psi }_{l}\left( \overrightarrow{r}\right) =\sum_{i}\psi
_{l_{i}}\left( \overrightarrow{v}_{i}\right) \delta \left( \overrightarrow{r}%
-\overrightarrow{q}_{i}\right) \delta _{l_{i}l}
\end{equation}%
and its average%
\begin{eqnarray}
\Psi _{l}\left( \overrightarrow{r},t\right) &=&\left\langle \widehat{\Psi }%
_{l}\left( \overrightarrow{r}\right) ;t\right\rangle =\frac{N}{V}%
\sum_{l_{1}}\int dx_{1}\;f_{l_{1}}\left( x_{1},t\right) \psi _{l_{i}}\left( 
\overrightarrow{v}_{1}\right) \delta \left( r-\overrightarrow{q}_{1}\right)
\delta _{l_{1}l} \\
&=&n\int d\overrightarrow{v}_{1}\;f_{l}\left( \overrightarrow{r},%
\overrightarrow{v}_{1},t\right) \psi _{l}\left( \overrightarrow{v}_{1}\right)
\notag
\end{eqnarray}%
The balance equation for this follows from the first BBGKY equation 
\begin{equation}
\left( \frac{d}{dt}+\overrightarrow{v}_{1}\cdot \frac{\partial }{\partial 
\overrightarrow{q}_{1}}\right) f_{l_{1}}\left( x_{1}\right)
=\sum_{l_{2}}\int dx_{2}\overline{T}_{-}\left( 12\right)
f_{l_{1}l_{2}}\left( x_{1}x_{2}\right)
\end{equation}%
and is%
\begin{eqnarray}
&&\frac{d}{dt}\Psi _{l}\left( \overrightarrow{r},t\right) +\overrightarrow{%
\nabla }\cdot \int d\overrightarrow{v}_{1}\;f_{l}\left( \overrightarrow{r},%
\overrightarrow{v}_{1},t\right) \overrightarrow{v}_{1}\psi _{l}\left( 
\overrightarrow{v}_{1}\right) \\
&=&\sum_{l_{1}l_{2}}\int dx_{1}\;\delta _{ll_{1}}\psi _{l_{1}}\left( 
\overrightarrow{v}_{1}\right) \delta \left( r-\overrightarrow{q}_{1}\right)
\int dx_{2}\overline{T}_{-}\left( 12\right) f_{ll_{2}}\left(
x_{1}x_{2}\right) .  \notag
\end{eqnarray}%
Introducing the specific velocity, $\overrightarrow{V}_{1}\left( 
\overrightarrow{r},t\right) =\overrightarrow{v}_{1}-\overrightarrow{u}\left( 
\overrightarrow{r},t\right) $, the second term on the left becomes%
\begin{eqnarray}
&&\int d\overrightarrow{v}_{1}\;f_{l}\left( \overrightarrow{r},%
\overrightarrow{v}_{1},t\right) \overrightarrow{v}_{1}\psi _{l}\left( 
\overrightarrow{v}_{1}\right) \\
&=&\overrightarrow{u}\left( \overrightarrow{r},t\right) \Psi _{l}\left( 
\overrightarrow{r},t\right) +\int d\overrightarrow{v}_{1}\;f_{l}\left( 
\overrightarrow{r},\overrightarrow{v}_{1},t\right) \overrightarrow{V}%
_{1}\left( \overrightarrow{r},t\right) \psi _{l}\left( \overrightarrow{v}%
_{1}\right)  \notag
\end{eqnarray}%
while it proves more convenient to rewrite the right hand side in terms of
the $\widehat{T}_{+}$ collision operator 
\begin{eqnarray}
&&\sum_{l_{1}l_{2}}\int dx_{1}\;\delta _{ll_{1}}\psi _{l_{1}}\left( 
\overrightarrow{v}_{1}\right) \delta \left( r-\overrightarrow{q}_{1}\right)
\int dx_{2}\overline{T}_{-}\left( 12\right) f_{ll_{2}}\left(
x_{1}x_{2}\right) \\
&=&\sum_{l_{1}l_{2}}\int dx_{1}dx_{2}f_{ll_{2}}\left( x_{1}x_{2}\right)
T_{+}\left( 12\right) \delta _{ll_{1}}\psi _{l_{1}}\left( \overrightarrow{v}%
_{1}\right) \delta \left( r-\overrightarrow{q}_{1}\right)  \notag
\end{eqnarray}%
so that the balance equation becomes%
\begin{eqnarray}
&&\frac{d}{dt}\Psi _{l}\left( \overrightarrow{r},t\right) +\overrightarrow{%
\nabla }\cdot \overrightarrow{u}\left( \overrightarrow{r},t\right) \Psi
_{l}\left( \overrightarrow{r},t\right) +\overrightarrow{\nabla }\cdot \int d%
\overrightarrow{v}_{1}\;f_{l}\left( \overrightarrow{r},\overrightarrow{v}%
_{1},t\right) \overrightarrow{V}_{1}\psi _{l}\left( \overrightarrow{v}%
_{1}\right) \\
&=&\sum_{l_{1}l_{2}}\int dx_{1}dx_{2}f_{l_{1}l_{2}}\left( x_{1}x_{2}\right)
\delta \left( r-\overrightarrow{q}_{1}\right) T_{+}\left( 12\right) \delta
_{ll_{1}}\psi _{l_{1}}\left( \overrightarrow{v}_{1}\right)  \notag
\end{eqnarray}%
with%
\begin{eqnarray}
&&T_{+}\left( 12\right) \delta _{ll_{1}}\psi _{l_{1}}\left( \overrightarrow{v%
}_{1}\right) \\
&=&-\overrightarrow{q}_{12}\cdot \overrightarrow{v}_{12}\delta \left(
q_{12}-\sigma _{l_{1}l_{2}}\right) \Theta \left( -\widehat{q}_{12}\cdot 
\overrightarrow{v}_{12}\right) \left( \sum_{l_{1}^{\prime }l_{2}^{\prime
}}K_{l_{1}l_{2}}^{l_{1}^{\prime }l_{2}^{\prime }}\left( x_{12}\right)
b_{l_{1}l_{2}}^{l_{1}^{\prime }l_{2}^{\prime }}-1\right) \delta
_{ll_{1}}\psi _{l_{1}}\left( \overrightarrow{v}_{1}\right)  \notag \\
&=&-\overrightarrow{q}_{12}\cdot \overrightarrow{v}_{12}\delta \left(
q_{12}-\sigma _{l_{1}l_{2}}\right) \Theta \left( -\widehat{q}_{12}\cdot 
\overrightarrow{v}_{12}\right) \sum_{l_{1}^{\prime }l_{2}^{\prime }}\left(
K_{l_{1}l_{2}}^{l_{1}^{\prime }l_{2}^{\prime }}\left( x_{12}\right)
b_{l_{1}l_{2}}^{l_{1}^{\prime }l_{2}^{\prime }}-\delta _{l_{1}l_{1}^{\prime
}}\delta _{l_{2}l_{2}^{\prime }}\right) \delta _{ll_{1}}\psi _{l_{1}}\left( 
\overrightarrow{v}_{1}\right) .  \notag
\end{eqnarray}%
In general, the right hand side can be separated into a sum of a flux and a
source term. Let%
\begin{equation}
B_{l_{1}l_{2},l_{1}^{\prime }l_{2}^{\prime };l}\left( x_{1},x_{2}\right)
=\left( K_{l_{1}l_{2}}^{l_{1}^{\prime }l_{2}^{\prime }}\left( x_{12}\right)
b_{l_{1}l_{2}}^{l_{1}^{\prime }l_{2}^{\prime }}-\delta _{l_{1}l_{1}^{\prime
}}\delta _{l_{2}l_{2}^{\prime }}\right) \delta _{ll_{1}}\psi _{l_{1}}\left( 
\overrightarrow{v}_{1}\right)
\end{equation}%
and define its even and odd components as%
\begin{eqnarray}
F_{l_{1}l_{2},l_{1}^{\prime }l_{2}^{\prime };l}\left( x_{1},x_{2}\right) &=&%
\frac{1}{2}\left( B_{l_{1}l_{2},l_{1}^{\prime }l_{2}^{\prime };l}\left(
x_{1},x_{2}\right) -B_{l_{1}l_{2},l_{1}^{\prime }l_{2}^{\prime };l}\left(
x_{2},x_{1}\right) \right) \\
&=&\frac{1}{2}\left( K_{l_{1}l_{2}}^{l_{1}^{\prime }l_{2}^{\prime }}\left(
x_{12}\right) b_{l_{1}l_{2}}^{l_{1}^{\prime }l_{2}^{\prime }}-\delta
_{l_{1}l_{1}^{\prime }}\delta _{l_{2}l_{2}^{\prime }}\right) \left( \delta
_{ll_{1}}\psi _{l_{1}}\left( \overrightarrow{v}_{1}\right) -\delta
_{ll_{2}}\psi _{l_{2}}\left( \overrightarrow{v}_{2}\right) \right)  \notag \\
S_{l_{1}l_{2},l_{1}^{\prime }l_{2}^{\prime };l}\left( x_{1},x_{2}\right) &=&%
\frac{1}{2}\left( B_{l_{1}l_{2},l_{1}^{\prime }l_{2}^{\prime };l}\left(
x_{1},x_{2}\right) +B_{l_{1}l_{2},l_{1}^{\prime }l_{2}^{\prime };l}\left(
x_{2},x_{1}\right) \right)  \notag \\
&=&\frac{1}{2}\left( K_{l_{1}l_{2}}^{l_{1}^{\prime }l_{2}^{\prime }}\left(
x_{12}\right) b_{l_{1}l_{2}}^{l_{1}^{\prime }l_{2}^{\prime }}-\delta
_{l_{1}l_{1}^{\prime }}\delta _{l_{2}l_{2}^{\prime }}\right) \left( \delta
_{ll_{1}}\psi _{l_{1}}\left( \overrightarrow{v}_{1}\right) +\delta
_{ll_{2}}\psi _{l_{2}}\left( \overrightarrow{v}_{2}\right) \right)  \notag
\end{eqnarray}%
so that 
\begin{eqnarray}
&&\frac{d}{dt}\Psi _{l}\left( \overrightarrow{r},t\right) +\overrightarrow{%
\nabla }\cdot \overrightarrow{u}\left( \overrightarrow{r},t\right) \Psi
_{l}\left( \overrightarrow{r},t\right) \\
&&+\overrightarrow{\nabla }\cdot \sum_{l}\int d\overrightarrow{v}%
_{1}\;f_{l}\left( \overrightarrow{r},\overrightarrow{v}_{1},t\right) 
\overrightarrow{V}_{1}\psi _{l}\left( \overrightarrow{v}_{1}\right)  \notag
\\
&=&-\sum_{l_{1}l_{2}l_{1}^{\prime }l_{2}^{\prime }}\int
dx_{1}dx_{2}f_{l_{1}l_{2}}\left( x_{1}x_{2}\right) \delta \left( 
\overrightarrow{r}-\overrightarrow{q}_{1}\right) \overrightarrow{q}%
_{12}\cdot \overrightarrow{v}_{12}\delta \left( q_{12}-\sigma
_{l_{1}l_{2}}\right) \Theta \left( -\widehat{q}_{12}\cdot \overrightarrow{v}%
_{12}\right)  \notag \\
&&\times \left( F_{l_{1}l_{2},l_{1}^{\prime }l_{2}^{\prime };l}\left(
x_{1},x_{2}\right) +S_{l_{1}l_{2},l_{1}^{\prime }l_{2}^{\prime };l}\left(
x_{1},x_{2}\right) \right) .  \notag
\end{eqnarray}%
Then, relabel the dummy variables to give%
\begin{eqnarray}
&&\sum_{l_{1}l_{2}l_{1}^{\prime }l_{2}^{\prime }}\int
dx_{1}dx_{2}f_{l_{1}l_{2}}\left( x_{1}x_{2}\right) \delta \left( 
\overrightarrow{r}-\overrightarrow{q}_{1}\right) \overrightarrow{q}%
_{12}\cdot \overrightarrow{v}_{12}\delta \left( q_{12}-\sigma
_{l_{1}l_{2}}\right) \\
&&\times \Theta \left( -\widehat{q}_{12}\cdot \overrightarrow{v}_{12}\right)
F_{l_{1}l_{2},l_{1}^{\prime }l_{2}^{\prime };l}\left( x_{1},x_{2}\right) 
\notag \\
&=&\frac{1}{2}\sum_{l_{1}l_{2}l_{1}^{\prime }l_{2}^{\prime }}\int
dx_{1}dx_{2}\overrightarrow{q}_{12}\cdot \overrightarrow{v}_{12}\delta
\left( q_{12}-\sigma _{l_{1}l_{2}}\right) \Theta \left( -\widehat{q}%
_{12}\cdot \overrightarrow{v}_{12}\right) F_{l_{1}l_{2},l_{1}^{\prime
}l_{2}^{\prime };l}\left( x_{1},x_{2}\right)  \notag \\
&&\times f_{l_{1}l_{2}}\left( x_{1}x_{2}\right) \left[ \delta \left( 
\overrightarrow{r}-\overrightarrow{q}_{1}\right) -\delta \left( 
\overrightarrow{r}-\overrightarrow{q}_{2}\right) \right]  \notag
\end{eqnarray}%
where use has been made of the the asymmetry of $F_{l_{1}l_{2},l_{1}^{\prime
}l_{2}^{\prime };l}\left( x_{1},x_{2}\right) $ and of the symmetry of the
distribution under an interchange of atoms. Finally, write%
\begin{eqnarray}
\delta \left( \overrightarrow{r}-\overrightarrow{q}_{1}\right) -\delta
\left( \overrightarrow{r}-\overrightarrow{q}_{2}\right) &=&\int_{0}^{1}dx\;%
\frac{d}{dx}\delta \left( \overrightarrow{r}-x\overrightarrow{q}_{1}-\left(
1-x\right) \overrightarrow{q}_{2}\right) \\
&=&-\overrightarrow{\nabla }\cdot \overrightarrow{q}_{12}\int_{0}^{1}dx\;%
\delta \left( \overrightarrow{r}-x\overrightarrow{q}_{1}-\left( 1-x\right) 
\overrightarrow{q}_{2}\right)  \notag
\end{eqnarray}%
so that the balance equation becomes%
\begin{equation}
\frac{d}{dt}\Psi _{l}\left( \overrightarrow{r},t\right) +\overrightarrow{%
\nabla }\cdot \overrightarrow{u}\left( \overrightarrow{r},t\right) \Psi
_{l}\left( \overrightarrow{r},t\right) +\overrightarrow{\nabla }\cdot 
\overrightarrow{F}_{l}\left( \overrightarrow{r},t\right) =S_{l}\left( 
\overrightarrow{r},t\right)  \label{Balance}
\end{equation}%
with the flux written as $\overrightarrow{F}_{l}\left( \overrightarrow{r}%
,t\right) =\overrightarrow{F}_{l}^{K}\left( \overrightarrow{r},t\right) +%
\overrightarrow{F}_{l}^{V}\left( \overrightarrow{r},t\right) $ where the
kinetic contribution is 
\begin{equation}
\overrightarrow{F}^{K}\left( \overrightarrow{r},t\right) =\int d%
\overrightarrow{v}_{1}\;f_{l}\left( \overrightarrow{r},\overrightarrow{v}%
_{1},t\right) \overrightarrow{V}_{1}\psi _{l}\left( \overrightarrow{v}%
_{1}\right)
\end{equation}%
and the collisional contribution is%
\begin{eqnarray}
\overrightarrow{F}^{V}\left( \overrightarrow{r},t\right)
&=&-\sum_{l_{1}l_{2}l_{1}^{\prime }l_{2}^{\prime }}\int dx_{1}dx_{2}\;%
\overrightarrow{q}_{12}\left( \overrightarrow{q}_{12}\cdot \overrightarrow{v}%
_{12}\right) \delta \left( q_{12}-\sigma _{l_{1}l_{2}}\right) \Theta \left( -%
\widehat{q}_{12}\cdot \overrightarrow{v}_{12}\right) \\
&&\times f_{l_{1}l_{2}}\left( x_{1}x_{2}\right) F_{l_{1}l_{2},l_{1}^{\prime
}l_{2}^{\prime };l}\left( x_{1},x_{2}\right) \int_{0}^{1}dx\;\delta \left( 
\overrightarrow{r}-x\overrightarrow{q}_{1}-\left( 1-x\right) \overrightarrow{%
q}_{2}\right)  \notag
\end{eqnarray}%
and the source is%
\begin{eqnarray}
S_{l}\left( \overrightarrow{r},t\right) &=&-\sum_{l_{1}l_{2}l_{1}^{\prime
}l_{2}^{\prime }}\int dx_{1}dx_{2}\;\left( \overrightarrow{q}_{12}\cdot 
\overrightarrow{v}_{12}\right) \delta \left( q_{12}-\sigma
_{l_{1}l_{2}}\right) \Theta \left( -\widehat{q}_{12}\cdot \overrightarrow{v}%
_{12}\right) \\
&&\times f_{l_{1}l_{2}}\left( x_{1}x_{2}\right) S_{l_{1}l_{2},l_{1}^{\prime
}l_{2}^{\prime };l}\left( x_{1},x_{2}\right) \delta \left( \overrightarrow{r}%
-\overrightarrow{q}_{1}\right) .  \notag
\end{eqnarray}

\subsection{Local number density}

Setting $\psi _{l}\left( \overrightarrow{v}_{1}\right) =1$ one has that%
\begin{eqnarray}
F_{l_{1}l_{2},l_{1}^{\prime }l_{2}^{\prime };l}\left( x_{1},x_{2}\right) &=&%
\frac{1}{2}\left( K_{l_{1}l_{2}}^{l_{1}^{\prime }l_{2}^{\prime }}\left(
x_{12}\right) \left( \delta _{ll_{1}^{\prime }}-\delta _{ll_{2}^{\prime
}}\right) -\delta _{l_{1}l_{1}^{\prime }}\delta _{l_{2}l_{2}^{\prime
}}\left( \delta _{ll_{1}}-\delta _{ll_{2}}\right) \right) \\
S_{l_{1}l_{2},l_{1}^{\prime }l_{2}^{\prime };l}\left( x_{1},x_{2}\right) &=&%
\frac{1}{2}\left( K_{l_{1}l_{2}}^{l_{1}^{\prime }l_{2}^{\prime }}\left(
x_{12}\right) \left( \delta _{ll_{1}^{\prime }}+\delta _{ll_{2}^{\prime
}}\right) -\delta _{l_{1}l_{1}^{\prime }}\delta _{l_{2}l_{2}^{\prime
}}\left( \delta _{ll_{1}}+\delta _{ll_{2}}\right) \right)  \notag
\end{eqnarray}%
From the normalization condition 
\begin{equation}
1=\sum_{lb}K_{l_{1}l_{2}}^{lb}
\end{equation}%
one has that%
\begin{eqnarray}
\sum_{l_{1}^{\prime }l_{2}^{\prime }}F_{l_{1}l_{2},l_{1}^{\prime
}l_{2}^{\prime };l}\left( x_{1},x_{2}\right) &=&\frac{1}{2}%
\sum_{l_{1}^{\prime }l_{2}^{\prime }}K_{l_{1}l_{2}}^{l_{1}^{\prime
}l_{2}^{\prime }}\left( x_{12}\right) \left( \delta _{ll_{1}^{\prime
}}-\delta _{ll_{2}^{\prime }}-\delta _{ll_{1}}+\delta _{ll_{2}}\right) \\
\sum_{l_{1}^{\prime }l_{2}^{\prime }}S_{l_{1}l_{2},l_{1}^{\prime
}l_{2}^{\prime };l}\left( x_{1},x_{2}\right) &=&\frac{1}{2}%
\sum_{l_{1}^{\prime }l_{2}^{\prime }}K_{l_{1}l_{2}}^{l_{1}^{\prime
}l_{2}^{\prime }}\left( x_{12}\right) \left( \delta _{ll_{1}^{\prime
}}-\delta _{ll_{2}^{\prime }}+\delta _{ll_{1}}-\delta _{ll_{2}}\right) 
\notag
\end{eqnarray}%
so that the balance equation becomes%
\begin{equation}
\frac{d}{dt}n_{l}+\overrightarrow{\nabla }\cdot \left( \overrightarrow{u}%
n_{l}\right) +\overrightarrow{\nabla }\cdot \overrightarrow{j}%
_{l}=S_{l}^{(n)}
\end{equation}%
with the source%
\begin{eqnarray}
S_{l}^{(n)} &=&-\frac{1}{2}\sum_{abl_{1}l_{2}}\int dx_{1}dx_{2}\;\left( 
\overrightarrow{q}_{12}\cdot \overrightarrow{v}_{12}\right) \delta \left(
q_{12}-\sigma _{l_{1}l_{2}}\right) \Theta \left( -\widehat{q}_{12}\cdot 
\overrightarrow{v}_{12}\right) \\
&&\times f_{l_{1}l_{2}}\left( x_{1}x_{2}\right) \delta \left( 
\overrightarrow{r}-\overrightarrow{q}_{1}\right) K_{l_{1}l_{2}}^{ab}\left(
x_{12}\right) \left( \delta _{al}+\delta _{bl}-\delta _{ll_{1}}-\delta
_{ll_{2}}\right)  \notag
\end{eqnarray}%
and the number current $\overrightarrow{j}_{l}=\overrightarrow{j}_{l}^{K}+%
\overrightarrow{j}_{l}^{V}$ with%
\begin{equation}
\overrightarrow{j}_{l}^{K}=\int d\overrightarrow{v}_{1}\;f_{l}\left( 
\overrightarrow{r},\overrightarrow{v}_{1},t\right) \overrightarrow{V}_{1}.
\end{equation}%
and%
\begin{eqnarray}
\overrightarrow{j}_{l}^{V} &=&-\frac{1}{2}\sum_{abl_{1}l_{2}}\int
dx_{1}dx_{2}\;\overrightarrow{q}_{12}\left( \overrightarrow{q}_{12}\cdot 
\overrightarrow{v}_{12}\right) \delta \left( q_{12}-\sigma
_{l_{1}l_{2}}\right) \Theta \left( -\widehat{q}_{12}\cdot \overrightarrow{v}%
_{12}\right) \\
&&\times f_{l_{1}l_{2}}\left( x_{1}x_{2}\right) K_{l_{1}l_{2}}^{ab}\left(
x_{12}\right) \left( \delta _{al}-\delta _{bl}-\delta _{ll_{1}}+\delta
_{ll_{2}}\right) \int_{0}^{1}dx\;\delta \left( \overrightarrow{r}-x%
\overrightarrow{q}_{1}-\left( 1-x\right) \overrightarrow{q}_{2}\right) . 
\notag
\end{eqnarray}

The balance equations for total number and mass density follow immediately.
Summing over $l$ gives%
\begin{equation}
\frac{d}{dt}n\left( \overrightarrow{r},t\right) +\overrightarrow{\nabla }%
\cdot \overrightarrow{u}\left( \overrightarrow{r},t\right) n\left( 
\overrightarrow{r},t\right) +\overrightarrow{\nabla }\cdot \sum_{l}%
\overrightarrow{j}_{l}^{K}=0
\end{equation}%
since the sum of the collisional contributions to the number current
vanishes. Similarly, multiplying by $m_{l}$ and then summing gives the
balance equation for local mass density%
\begin{equation}
\frac{d}{dt}\rho \left( \overrightarrow{r},t\right) +\overrightarrow{\nabla }%
\cdot \overrightarrow{u}\left( \overrightarrow{r},t\right) \rho \left( 
\overrightarrow{r},t\right) +\overrightarrow{\nabla }\cdot \overrightarrow{Q}%
=S^{(\rho )}.
\end{equation}%
Here, the anomalous mass flux is 
\begin{eqnarray}
\overrightarrow{Q} &=&-\frac{1}{2}\sum_{abl_{1}l_{2}}\int dx_{1}dx_{2}\;%
\overrightarrow{q}_{12}\left( \overrightarrow{q}_{12}\cdot \overrightarrow{v}%
_{12}\right) \delta \left( q_{12}-\sigma _{l_{1}l_{2}}\right) \Theta \left( -%
\widehat{q}_{12}\cdot \overrightarrow{v}_{12}\right) \\
&&\times f_{l_{1}l_{2}}\left( x_{1}x_{2}\right) K_{l_{1}l_{2}}^{ab}\left(
x_{12}\right) \left( m_{a}-m_{b}-m_{l_{1}}+m_{l_{2}}\right)
\int_{0}^{1}dx\;\delta \left( \overrightarrow{r}-x\overrightarrow{q}%
_{1}-\left( 1-x\right) \overrightarrow{q}_{2}\right)  \notag
\end{eqnarray}%
and the mass source term is%
\begin{eqnarray}
S_{l}^{(\rho )} &=&\frac{1}{2}\sum_{abl_{1}l_{2}}\int dx_{1}dx_{2}\;\left( 
\overrightarrow{q}_{12}\cdot \overrightarrow{v}_{12}\right) \delta \left(
q_{12}-\sigma _{l_{1}l_{2}}\right) \Theta \left( -\widehat{q}_{12}\cdot 
\overrightarrow{v}_{12}\right) \\
&&\times f_{l_{1}l_{2}}\left( x_{1}x_{2}\right) \delta \left( 
\overrightarrow{r}-\overrightarrow{q}_{1}\right) K_{l_{1}l_{2}}^{ab}\left(
x_{12}\right) \delta m_{l_{1}l_{2}}^{ab},  \notag
\end{eqnarray}%
which is only nonzero if the collisions do not conserve mass.

\subsection{Momentum density}

Taking $\psi _{l}\left( \overrightarrow{v}_{1}\right) =m_{l}\overrightarrow{v%
}_{1}=\overrightarrow{p}_{1}$ in Eq.(\ref{Balance}) and summing over $l$
gives%
\begin{equation}
\frac{\partial }{\partial t}\rho \overrightarrow{u}+\overrightarrow{\nabla }%
\cdot \left( \rho \overrightarrow{u}\overrightarrow{u}\right) +%
\overrightarrow{\nabla }\cdot \left( \overleftrightarrow{P}+\overrightarrow{Q%
}\overrightarrow{u}\right) =\overrightarrow{S}^{(p)}
\end{equation}%
with $\overleftrightarrow{P}=\overleftrightarrow{P}^{K}+\overleftrightarrow{P%
}^{V}+\overleftrightarrow{P}^{M}$ where the kinetic contribution is 
\begin{equation}
\overleftrightarrow{P}^{K}=\sum_{l}m_{l}\int d\overrightarrow{v}%
_{1}\;f_{l}\left( \overrightarrow{r},\overrightarrow{v}_{1},t\right) 
\overrightarrow{V}_{1}\overrightarrow{V}_{1}.
\end{equation}%
To explicitly write the remaining flux and source terms, we need%
\begin{equation}
\left( b_{l_{1}l_{2}}^{l_{1}^{\prime }l_{2}^{\prime }}\overrightarrow{p}_{1}-%
\overrightarrow{p}_{1}\right) =\frac{1}{2}\left( \overrightarrow{\gamma }%
_{l_{1}l_{2}}^{l_{1}^{\prime }l_{2}^{\prime }}-\delta
m_{l_{1}l_{2}}^{l_{1}^{\prime }l_{2}^{\prime }}\overrightarrow{V}_{12}\right)
\end{equation}%
giving%
\begin{eqnarray}
S_{l_{1}l_{2}}^{l_{1}^{\prime }l_{2}^{\prime }} &=&-\frac{1}{2}\delta
m_{l_{1}l_{2}}^{l_{1}^{\prime }l_{2}^{\prime }}\left( \overrightarrow{V}%
_{12}-\overrightarrow{u}\right) -\frac{1}{2}\delta
m_{l_{1}l_{2}}^{l_{1}^{\prime }l_{2}^{\prime }}\overrightarrow{u} \\
F_{l_{1}l_{2}}^{l_{1}^{\prime }l_{2}^{\prime }} &=&\frac{1}{2}%
\overrightarrow{\gamma }_{l_{1}l_{2}}^{l_{1}^{\prime }l_{2}^{\prime }}. 
\notag
\end{eqnarray}%
At this point, it is useful to separate $\overrightarrow{\gamma }%
_{l_{1}l_{2}}^{l_{1}^{\prime }l_{2}^{\prime }}$ into two parts: its value in
the local rest-frame of the colliding atoms, $\widetilde{\overrightarrow{%
\gamma }}_{l_{1}l_{2}}^{l_{1}^{\prime }l_{2}^{\prime }}$, and the part
coming from the Galilean transformation to the lab frame%
\begin{equation*}
\overrightarrow{\gamma }_{l_{1}l_{2}}^{l_{1}^{\prime }l_{2}^{\prime }}=%
\widetilde{\overrightarrow{\gamma }}_{l_{1}l_{2}}^{l_{1}^{\prime
}l_{2}^{\prime }}+\left( m_{l_{1}^{\prime }}-m_{l_{1}}-m_{l_{2}^{\prime
}}-m_{l_{2}}\right) \overrightarrow{V}_{12}
\end{equation*}%
so that we have that the collisional part of the flux is%
\begin{eqnarray}
\overleftrightarrow{P}^{V} &=&-\frac{1}{2}\sum_{l_{1}l_{2}l_{1}^{\prime
}l_{2}^{\prime }}\int dx_{1}dx_{2}\;\overrightarrow{q}_{12}\left( 
\overrightarrow{q}_{12}\cdot \overrightarrow{v}_{12}\right) \delta \left(
q_{12}-\sigma _{l_{1}l_{2}}\right) \Theta \left( -\widehat{q}_{12}\cdot 
\overrightarrow{v}_{12}\right) \\
&&\times f_{l_{1}l_{2}}\left( x_{1}x_{2}\right)
K_{l_{1}l_{2}}^{l_{1}^{\prime }l_{2}^{\prime }}\left( x_{12}\right) 
\widetilde{\overrightarrow{\gamma }}_{l_{1}l_{2}}^{l_{1}^{\prime
}l_{2}^{\prime }}\int_{0}^{1}dx\;\delta \left( \overrightarrow{r}-x%
\overrightarrow{q}_{1}-\left( 1-x\right) \overrightarrow{q}_{2}\right) 
\notag
\end{eqnarray}%
and the contribution from the instantaneous exchange of mass is%
\begin{eqnarray}
\overleftrightarrow{P}^{M} &=&-\frac{1}{2}\sum_{l_{1}l_{2}l_{1}^{\prime
}l_{2}^{\prime }}\int dx_{1}dx_{2}\;\overrightarrow{q}_{12}\left( 
\overrightarrow{q}_{12}\cdot \overrightarrow{v}_{12}\right) \delta \left(
q_{12}-\sigma _{l_{1}l_{2}}\right) \Theta \left( -\widehat{q}_{12}\cdot 
\overrightarrow{v}_{12}\right) \\
&&\times f_{l_{1}l_{2}}\left( x_{1}x_{2}\right)
K_{l_{1}l_{2}}^{l_{1}^{\prime }l_{2}^{\prime }}\left( x_{12}\right) \left( 
\overrightarrow{V}_{12}-\overrightarrow{u}\right) \left( m_{l_{1}^{\prime
}}-m_{l_{1}}-m_{l_{2}^{\prime }}-m_{l_{2}}\right) \int_{0}^{1}dx\;\delta
\left( \overrightarrow{r}-x\overrightarrow{q}_{1}-\left( 1-x\right) 
\overrightarrow{q}_{2}\right)  \notag
\end{eqnarray}%
and the source can be written as%
\begin{eqnarray}
\overrightarrow{S}^{(p)} &=&\overrightarrow{u}S^{(\rho )}+\overline{%
\overrightarrow{S}}^{(p)} \\
\overline{\overrightarrow{S}}^{(p)} &=&\frac{1}{2}\sum_{l_{1}l_{2}l_{1}^{%
\prime }l_{2}^{\prime }}\int dx_{1}dx_{2}\;\left( \overrightarrow{q}%
_{12}\cdot \overrightarrow{v}_{12}\right) \delta \left( q_{12}-\sigma
_{l_{1}l_{2}}\right) \Theta \left( -\widehat{q}_{12}\cdot \overrightarrow{v}%
_{12}\right)  \notag \\
&&\times \left( \overrightarrow{V}-\overrightarrow{u}\right)
f_{l_{1}l_{2}}\left( x_{1}x_{2}\right) K_{l_{1}l_{2}}^{l_{1}^{\prime
}l_{2}^{\prime }}\left( x_{12}\right) \delta m_{l_{1}l_{2}}^{l_{1}^{\prime
}l_{2}^{\prime }}\delta \left( \overrightarrow{r}-\overrightarrow{q}%
_{1}\right)  \notag
\end{eqnarray}%
By using the balance equation for total mass density%
\begin{equation}
\frac{d}{dt}\rho \overrightarrow{u}+\overrightarrow{\nabla }\cdot \left( 
\overrightarrow{u}\rho \overrightarrow{u}\right) =\rho \frac{d}{dt}%
\overrightarrow{u}+\rho \overrightarrow{u}\cdot \overrightarrow{\nabla }%
\overrightarrow{u}-\overrightarrow{u}\overrightarrow{\nabla }\cdot 
\overrightarrow{Q}+\overrightarrow{u}S^{(\rho )}
\end{equation}%
we can write 
\begin{equation}
\frac{\partial }{\partial t}\overrightarrow{u}+\overrightarrow{u}\cdot 
\overrightarrow{\nabla }\overrightarrow{u}+\rho ^{-1}\left( \overrightarrow{%
\nabla }\cdot \overleftrightarrow{P}+\overrightarrow{Q}\cdot \overrightarrow{%
\nabla }\overrightarrow{u}\right) =\rho ^{-1}\overline{\overrightarrow{S}}%
^{(p)}\text{.}
\end{equation}

\subsection{Energy}

Taking $\psi _{l}\left( \overrightarrow{v}_{1}\right) =\frac{1}{2}%
m_{l}v_{1}^{2}$ in Eq.(\ref{Balance}) and summing over $l$ gives%
\begin{equation}
\frac{d}{dt}E+\overrightarrow{\nabla }\cdot \left( \overrightarrow{u}%
E\right) +\overrightarrow{\nabla }\cdot \overrightarrow{F}=S^{(E)}
\end{equation}%
where the kinetic part of the flux is%
\begin{eqnarray}
\overrightarrow{F}^{K} &=&\sum_{l}\frac{1}{2}m_{l}\int d\overrightarrow{v}%
_{1}\;f_{l}\left( \overrightarrow{r},\overrightarrow{v}_{1},t\right) 
\overrightarrow{V}_{1}v_{1}^{2}  \label{E2} \\
&=&\sum_{l}\frac{1}{2}m_{l}\int d\overrightarrow{v}_{1}\;f_{l}\left( 
\overrightarrow{r},\overrightarrow{v}_{1},t\right) \overrightarrow{V}%
_{1}\left( \overrightarrow{V}_{1}+\overrightarrow{u}\right) ^{2} \\
&=&\overrightarrow{q}^{K}+\overrightarrow{u}\cdot \overleftrightarrow{P}^{K}
\notag
\end{eqnarray}%
with the kinetic contribution to the heat flux being defined as%
\begin{equation}
\overrightarrow{q}^{K}\equiv \sum_{l}\frac{1}{2}m_{l}\int d\overrightarrow{v}%
_{1}\;f_{l}\left( \overrightarrow{r},\overrightarrow{v}_{1},t\right) 
\overrightarrow{V}_{1}V_{1}^{2}.
\end{equation}

The source term comes from the even part of the collision kernel%
\begin{eqnarray}
S_{l_{1}l_{2}}^{l_{1}^{\prime }l_{2}^{\prime }} &=&\frac{1}{2}\left(
b_{l_{1}l_{2}}^{l_{1}^{\prime }l_{2}^{\prime }}-1\right) \left( \frac{1}{%
2m_{l_{1}}}p_{1}^{2}+\frac{1}{2m_{l_{2}}}p_{2}^{2}\right) \\
&=&\frac{1}{2}\left[ \frac{1}{2m_{l_{1}^{\prime }}}p_{1}^{\prime 2}+\frac{1}{%
2m_{l_{2}^{\prime }}}p_{2}^{\prime 2}-\frac{1}{2m_{l_{1}}}p_{1}^{2}-\frac{1}{%
2m_{l_{2}}}p_{2}^{2}\right]  \notag \\
&=&-\frac{1}{2}\left[ \overline{\delta E_{l_{1}l_{2}}^{l_{1}^{\prime
}l_{2}^{\prime }}}+\frac{1}{2}\delta m_{l_{1}l_{2}}^{l_{1}^{\prime
}l_{2}^{\prime }}V^{2}\right]  \notag \\
&=&-\frac{1}{2}\left[ \overline{\delta E_{l_{1}l_{2}}^{l_{1}^{\prime
}l_{2}^{\prime }}}+\frac{1}{2}\delta m_{l_{1}l_{2}}^{l_{1}^{\prime
}l_{2}^{\prime }}\left( \overrightarrow{V}-\overrightarrow{u}\right) ^{2}%
\right] -\frac{1}{2}\delta m_{l_{1}l_{2}}^{l_{1}^{\prime }l_{2}^{\prime
}}\left( \overrightarrow{V}-\overrightarrow{u}\right) \cdot \overrightarrow{u%
}-\frac{1}{4}\delta m_{l_{1}l_{2}}^{l_{1}^{\prime }l_{2}^{\prime }}u^{2} 
\notag
\end{eqnarray}%
so that 
\begin{equation}
S^{(E)}=\xi +\overrightarrow{u}\cdot \overline{\overrightarrow{S}}^{(p)}+%
\frac{1}{2}u^{2}\overline{S}^{(\rho )}
\end{equation}%
with the (rest-frame) source term%
\begin{eqnarray}
\xi &=&\frac{1}{2}\sum_{l_{1}l_{2}l_{1}^{\prime }l_{2}^{\prime }}\int
dx_{1}dx_{2}\;\left( \overrightarrow{q}_{12}\cdot \overrightarrow{v}%
_{12}\right) \delta \left( q_{12}-\sigma _{l_{1}l_{2}}\right) \Theta \left( -%
\widehat{q}_{12}\cdot \overrightarrow{v}_{12}\right) \\
&&\times \left[ \overline{\delta E_{l_{1}l_{2}}^{l_{1}^{\prime
}l_{2}^{\prime }}}+\frac{1}{2}\delta m_{l_{1}l_{2}}^{l_{1}^{\prime
}l_{2}^{\prime }}\left( \overrightarrow{V}-\overrightarrow{u}\right) ^{2}%
\right] f_{l_{1}l_{2}}\left( x_{1}x_{2}\right) K_{l_{1}l_{2}}^{l_{1}^{\prime
}l_{2}^{\prime }}\left( x_{12}\right) \delta \left( \overrightarrow{r}-%
\overrightarrow{q}_{1}\right) .  \notag
\end{eqnarray}

The flux comes from the odd part of the collision kernel 
\begin{eqnarray}
F_{l_{1}l_{2}}^{l_{1}^{\prime }l_{2}^{\prime }} &=&\frac{1}{2}\left(
b_{l_{1}l_{2}}^{l_{1}^{\prime }l_{2}^{\prime }}-1\right) \left( \frac{1}{%
2m_{l_{1}}}p_{1}^{2}-\frac{1}{2m_{l_{2}}}p_{2}^{2}\right) \\
&=&\frac{1}{2}\left( b_{l_{1}l_{2}}^{l_{1}^{\prime }l_{2}^{\prime
}}-1\right) \left( \frac{1}{m_{l_{1}}+m_{l_{2}}}\left(
p_{1}^{2}-p_{2}^{2}\right) -\frac{m_{l_{1}}-m_{l_{2}}}{m_{l_{1}}+m_{l_{2}}}%
\left( \frac{1}{2m_{l_{1}}}p_{1}^{2}+\frac{1}{2m_{l_{2}}}p_{2}^{2}\right)
\right)  \notag
\end{eqnarray}%
The first term gives%
\begin{eqnarray}
&&\left( b_{l_{1}l_{2}}^{l_{1}^{\prime }l_{2}^{\prime }}-1\right) \frac{1}{%
m_{l_{1}}+m_{l_{2}}}\left( p_{1}^{2}-p_{2}^{2}\right) \\
&=&\frac{1}{m_{l_{1}^{\prime }}+m_{l_{2}^{\prime }}}\left[ \left( 
\overrightarrow{p}_{1}+\frac{1}{2}\overrightarrow{\gamma }%
_{l_{1}l_{2}}^{l_{1}^{\prime }l_{2}^{\prime }}-\frac{1}{2}\delta
m_{l_{1}l_{2}}^{l_{1}^{\prime }l_{2}^{\prime }}\overrightarrow{V}\right)
^{2}-\left( \overrightarrow{p}_{2}-\frac{1}{2}\overrightarrow{\gamma }%
_{l_{1}l_{2}}^{l_{1}^{\prime }l_{2}^{\prime }}-\frac{1}{2}\delta
m_{l_{1}l_{2}}^{l_{1}^{\prime }l_{2}^{\prime }}\overrightarrow{V}\right) ^{2}%
\right]  \notag \\
&&-\frac{1}{m_{l_{1}}+m_{l_{2}}}\left( p_{1}^{2}-p_{2}^{2}\right)  \notag \\
&=&\overrightarrow{V}\cdot \overrightarrow{\gamma }_{l_{1}l_{2}}^{l_{1}^{%
\prime }l_{2}^{\prime }}+\frac{\delta m_{l_{1}l_{2}}^{l_{1}^{\prime
}l_{2}^{\prime }}}{m_{l_{1}^{\prime }}+m_{l_{2}^{\prime }}}\overrightarrow{p}%
_{12}\cdot \overrightarrow{V}+\left( \frac{1}{m_{l_{1}^{\prime
}}+m_{l_{2}^{\prime }}}-\frac{1}{m_{l_{1}}+m_{l_{2}}}\right) \left(
p_{1}^{2}-p_{2}^{2}\right)  \notag
\end{eqnarray}%
while the second is 
\begin{eqnarray}
&&\left( b_{l_{1}l_{2}}^{l_{1}^{\prime }l_{2}^{\prime }}-1\right) \frac{%
m_{l_{1}}-m_{l_{2}}}{m_{l_{1}}+m_{l_{2}}}\left( \frac{1}{2m_{l_{1}}}%
p_{1}^{2}+\frac{1}{2m_{l_{2}}}p_{2}^{2}\right) \\
&=&-\frac{m_{l_{1}^{\prime }}-m_{l_{2}^{\prime }}}{m_{l_{1}^{\prime
}}+m_{l_{2}^{\prime }}}\left( \overline{\delta E_{l_{1}l_{2}}^{l_{1}^{\prime
}l_{2}^{\prime }}}+\frac{1}{2}\delta m_{l_{1}l_{2}}^{l_{1}^{\prime
}l_{2}^{\prime }}V^{2}\right) +\left( \frac{m_{l_{1}^{\prime
}}-m_{l_{2}^{\prime }}}{m_{l_{1}^{\prime }}+m_{l_{2}^{\prime }}}-\frac{%
m_{l_{1}}-m_{l_{2}}}{m_{l_{1}}+m_{l_{2}}}\right) \left( \frac{1}{2m_{l_{1}}}%
p_{1}^{2}+\frac{1}{2m_{l_{2}}}p_{2}^{2}\right)  \notag \\
&=&-\frac{m_{l_{1}^{\prime }}-m_{l_{2}^{\prime }}}{m_{l_{1}^{\prime
}}+m_{l_{2}^{\prime }}}\left( \overline{\delta E_{l_{1}l_{2}}^{l_{1}^{\prime
}l_{2}^{\prime }}}+\frac{1}{2}\delta m_{l_{1}l_{2}}^{l_{1}^{\prime
}l_{2}^{\prime }}V^{2}\right) +\left( \frac{1}{m_{l_{1}^{\prime
}}+m_{l_{2}^{\prime }}}\right) \left( \frac{m_{l_{1}^{\prime }}}{m_{l_{1}}}%
p_{1}^{2}-\frac{m_{l_{2}^{\prime }}}{m_{l_{2}}}p_{2}^{2}\right)  \notag \\
&&-\left( \frac{1}{m_{l_{1}}+m_{l_{2}}}\right) \left(
p_{1}^{2}-p_{2}^{2}\right)  \notag
\end{eqnarray}%
so%
\begin{eqnarray}
2F_{l_{1}l_{2}}^{l_{1}^{\prime }l_{2}^{\prime }} &=&\overrightarrow{V}%
_{12}\cdot \overrightarrow{\gamma }_{l_{1}l_{2}}^{l_{1}^{\prime
}l_{2}^{\prime }}+\frac{\delta m_{l_{1}l_{2}}^{l_{1}^{\prime }l_{2}^{\prime
}}}{m_{l_{1}^{\prime }}+m_{l_{2}^{\prime }}}\overrightarrow{p}_{12}\cdot 
\overrightarrow{V}_{12}+\frac{m_{l_{1}^{\prime }}-m_{l_{2}^{\prime }}}{%
m_{l_{1}^{\prime }}+m_{l_{2}^{\prime }}}\left( \overline{\delta
E_{l_{1}l_{2}}^{l_{1}^{\prime }l_{2}^{\prime }}}+\frac{1}{2}\delta
m_{l_{1}l_{2}}^{l_{1}^{\prime }l_{2}^{\prime }}V_{12}^{2}\right) \\
&&+\left( \frac{m_{l_{1}}-m_{l_{1}^{\prime }}}{m_{l_{1}^{\prime
}}+m_{l_{2}^{\prime }}}\right) \frac{1}{m_{l_{1}}}p_{1}^{2}+\left( \frac{%
m_{l_{2}^{\prime }}-m_{l_{2}}}{m_{l_{1}^{\prime }}+m_{l_{2}^{\prime }}}%
\right) \frac{1}{m_{l_{2}}}p_{2}^{2}  \notag
\end{eqnarray}%
which gives, after some algebra,%
\begin{eqnarray}
2F_{l_{1}l_{2}}^{l_{1}^{\prime }l_{2}^{\prime }} &=&\overrightarrow{V}%
_{12}\cdot \widetilde{\overrightarrow{\gamma }}_{l_{1}l_{2}}^{l_{1}^{\prime
}l_{2}^{\prime }}+\frac{1}{m_{l_{1}}+m_{l_{2}}}\left( \frac{%
m_{l_{1}}m_{l_{2}^{\prime }}-m_{l_{2}}m_{l_{1}^{\prime }}}{m_{l_{1}^{\prime
}}+m_{l_{2}^{\prime }}}\right) \mu _{l_{1}l_{2}}v^{2} \\
&&+\frac{m_{l_{1}^{\prime }}-m_{l_{2}^{\prime }}}{m_{l_{1}^{\prime
}}+m_{l_{2}^{\prime }}}\overline{\delta E_{l_{1}l_{2}}^{l_{1}^{\prime
}l_{2}^{\prime }}}-\frac{1}{2}\left( -m_{l_{1}^{\prime }}+m_{l_{2}^{\prime
}}-m_{l_{2}}+m_{l_{1}}\right) V_{12}^{2}  \notag
\end{eqnarray}%
This can also be written as 
\begin{eqnarray}
2F_{l_{1}l_{2}}^{l_{1}^{\prime }l_{2}^{\prime }} &=&\left( \overrightarrow{V}%
_{12}-\overrightarrow{u}\right) \cdot \widetilde{\overrightarrow{\gamma }}%
_{l_{1}l_{2}}^{l_{1}^{\prime }l_{2}^{\prime }}+\frac{m_{l_{2}^{\prime
}}m_{l_{1}}-m_{l_{1}^{\prime }}m_{l_{2}}}{\left( m_{l_{2}^{\prime
}}+m_{l_{1}^{\prime }}\right) \left( m_{l_{2}}+m_{l_{1}}\right) }\mu
_{l_{1}l_{2}}v^{2} \\
&&+\frac{m_{l_{1}^{\prime }}-m_{l_{2}^{\prime }}}{m_{l_{1}^{\prime
}}+m_{l_{2}^{\prime }}}\overline{\delta E_{l_{1}l_{2}}^{l_{1}^{\prime
}l_{2}^{\prime }}}-\frac{1}{2}\left( -m_{l_{1}^{\prime }}+m_{l_{2}^{\prime
}}-m_{l_{2}}+m_{l_{1}}\right) \left( \overrightarrow{V}_{12}-\overrightarrow{%
u}\right) ^{2}  \notag \\
&&+\left( m_{l_{1}^{\prime }}-m_{l_{1}}-m_{l_{2}^{\prime }}+m_{l_{2}}\right)
\left( \left( \overrightarrow{V}_{12}-\overrightarrow{u}\right) \cdot 
\overrightarrow{u}\right) +\frac{1}{2}\left( m_{l_{1}^{\prime
}}-m_{l_{1}}-m_{l_{2}^{\prime }}+m_{l_{2}}\right) u^{2} \\
&&+\overrightarrow{u}\cdot \widetilde{\overrightarrow{\gamma }}%
_{l_{1}l_{2}}^{l_{1}^{\prime }l_{2}^{\prime }}  \notag
\end{eqnarray}%
so that%
\begin{equation}
\overrightarrow{F}^{(V)}=\overrightarrow{q}^{V}+\overrightarrow{q}^{m}+%
\overrightarrow{q}^{\delta E}+\overrightarrow{u}.\left( \overleftrightarrow{P%
}^{V}+\overleftrightarrow{P}^{M}\right) +\frac{1}{2}u^{2}\overrightarrow{Q}
\end{equation}%
where the different pieces of the heat flux vector are the usual collisional
contribution%
\begin{eqnarray}
\overrightarrow{q}^{V} &=&-\frac{1}{2}\sum_{l_{1}l_{2}l_{1}^{\prime
}l_{2}^{\prime }}\int dx_{1}dx_{2}\;\overrightarrow{q}_{12}\left( 
\overrightarrow{q}_{12}\cdot \overrightarrow{v}_{12}\right) \delta \left(
q_{12}-\sigma _{l_{1}l_{2}}\right) \Theta \left( -\widehat{q}_{12}\cdot 
\overrightarrow{v}_{12}\right) \\
&&\times f_{l_{1}l_{2}}\left( x_{1}x_{2}\right)
K_{l_{1}l_{2}}^{l_{1}^{\prime }l_{2}^{\prime }}\left( x_{12}\right) \left( 
\overrightarrow{V}_{12}-\overrightarrow{u}\right) \cdot \widetilde{%
\overrightarrow{\gamma }}_{l_{1}l_{2}}^{l_{1}^{\prime }l_{2}^{\prime
}}\int_{0}^{1}dx\;\delta \left( \overrightarrow{r}-x\overrightarrow{q}%
_{1}-\left( 1-x\right) \overrightarrow{q}_{2}\right) ,  \notag
\end{eqnarray}%
a part arising from the instantaneous transfer of mass%
\begin{eqnarray}
\overrightarrow{q}^{m} &=&-\frac{1}{2}\sum_{l_{1}l_{2}l_{1}^{\prime
}l_{2}^{\prime }}\frac{m_{l_{2}^{\prime }}m_{l_{1}}-m_{l_{1}^{\prime
}}m_{l_{2}}}{\left( m_{l_{2}^{\prime }}+m_{l_{1}^{\prime }}\right) \left(
m_{l_{2}}+m_{l_{1}}\right) } \\
&&\times \int dx_{1}dx_{2}\;\overrightarrow{q}_{12}\left( \overrightarrow{q}%
_{12}\cdot \overrightarrow{v}_{12}\right) \delta \left( q_{12}-\sigma
_{l_{1}l_{2}}\right) \Theta \left( -\widehat{q}_{12}\cdot \overrightarrow{v}%
_{12}\right)  \notag \\
&&\times f_{l_{1}l_{2}}\left( x_{1}x_{2}\right)
K_{l_{1}l_{2}}^{l_{1}^{\prime }l_{2}^{\prime }}\left( x_{12}\right) \mu
_{l_{1}l_{2}}v^{2}  \notag \\
&&\times \int_{0}^{1}dx\;\delta \left( \overrightarrow{r}-x\overrightarrow{q}%
_{1}-\left( 1-x\right) \overrightarrow{q}_{2}\right) ,  \notag
\end{eqnarray}%
and a part arising from the loss of energy%
\begin{eqnarray}
\overrightarrow{q}^{\delta E} &=&-\frac{1}{2}\sum_{l_{1}l_{2}l_{1}^{\prime
}l_{2}^{\prime }}\frac{m_{l_{1}^{\prime }}-m_{l_{2}^{\prime }}}{%
m_{l_{1}^{\prime }}+m_{l_{2}^{\prime }}}\int dx_{1}dx_{2}\;\overrightarrow{q}%
_{12}\left( \overrightarrow{q}_{12}\cdot \overrightarrow{v}_{12}\right)
\delta \left( q_{12}-\sigma _{l_{1}l_{2}}\right) \Theta \left( -\widehat{q}%
_{12}\cdot \overrightarrow{v}_{12}\right) \\
&&\times f_{l_{1}l_{2}}\left( x_{1}x_{2}\right)
K_{l_{1}l_{2}}^{l_{1}^{\prime }l_{2}^{\prime }}\left( x_{12}\right) 
\overline{\delta E_{l_{1}l_{2}}^{l_{1}^{\prime }l_{2}^{\prime }}}%
\int_{0}^{1}dx\;\delta \left( \overrightarrow{r}-x\overrightarrow{q}%
_{1}-\left( 1-x\right) \overrightarrow{q}_{2}\right) .  \notag
\end{eqnarray}%
\bigskip A little rearrangement allows us to write the energy balance
equation as%
\begin{equation}
\frac{\partial }{\partial t}E+\overrightarrow{\nabla }\cdot \left( 
\overrightarrow{u}E\right) +\overrightarrow{\nabla }\cdot \overrightarrow{q}+%
\overrightarrow{\nabla }\cdot \left( \overrightarrow{u}\cdot 
\overleftrightarrow{P}\right) +\overrightarrow{\nabla }\cdot \left( \frac{1}{%
2}u^{2}\overrightarrow{Q}\right) =\xi +\overrightarrow{u}\cdot \overline{%
\overrightarrow{S}}^{(p)}+\frac{1}{2}u^{2}\overline{S}^{(\rho )}.
\end{equation}%
Alternatively, noting the relation between the total energy and the kinetic
temperature%
\begin{equation}
E=\frac{D}{2}nk_{B}T+\frac{1}{2}\rho u^{2}
\end{equation}%
gives an equation for the evolution of the kinetic temperature%
\begin{equation}
\left( \frac{\partial }{\partial t}+\overrightarrow{u}\cdot \overrightarrow{%
\nabla }\right) T-\frac{T}{n}\overrightarrow{\nabla }\cdot \sum_{l}%
\overrightarrow{j}_{l}^{K}+\frac{2}{Dnk_{B}}\left[ \overleftrightarrow{P}:%
\overrightarrow{\nabla }\overrightarrow{u}+\overrightarrow{\nabla }\cdot 
\overrightarrow{q}\right] =\xi .
\end{equation}

\bigskip

\bibliographystyle{apsrev}
\bibliography{physics}

\end{document}